\documentclass[preprintnumbers,nofootinbib]{revtex4-2}
\usepackage{graphicx}
\usepackage{dcolumn}
\usepackage{bm}
\usepackage{epsfig,amsmath}
\usepackage{amssymb}
\usepackage{subfigure}
\usepackage{subcaption}
\usepackage{float}
\usepackage{orcidlink}
\usepackage{url}
\usepackage{comment}
\usepackage{ulem}
 \definecolor{bluish}{rgb}{0.2,0.5,0.8}
 \definecolor{redish}{rgb}{0.7,0.2,0.0}  
\hypersetup{linkcolor=redish,          
	citecolor=blue,        
	filecolor=magenta,      
	urlcolor=bluish} 

\DeclareFontFamily{U}{rsfs}{}         
\DeclareFontShape{U}{rsfs}{m}{n}{<5> rsfs5 <6><7> rsfs7          %
  <8><9><10><10.95><12><14.4><17.28><20.74><24.88> rsfs10}{}     %
\DeclareMathAlphabet{\mathfs}{U}{rsfs}{m}{n}

\begin{document}

\title{Orbital dynamics and precession in magnetized Kerr spacetime}
\author{Karthik Iyer \orcidlink{0009-0000-9362-6771}}
\email{karthik.mcnsmpl2023@learner.manipal.edu}
\affiliation{Manipal Centre for Natural Sciences, Manipal Academy of Higher Education, Manipal 576104, India}
\author{Chandrachur Chakraborty \orcidlink{0000-0003-4380-3033
}}
\email{chandrachur.c@manipal.edu} 
\affiliation{Manipal Centre for Natural Sciences, Manipal Academy of Higher Education, Manipal 576104, India}

\begin{abstract}
We study the orbital structure and precession dynamics of neutral test particles in the magnetized Kerr black hole (MKBH) spacetime—an exact electrovacuum solution of the Einstein–Maxwell equations that self-consistently incorporates the curvature effects of an external magnetic field. This geometry allows a unified treatment of gravitational and magnetic influences across weak to ultra-strong regimes. The analysis reveals a critical magnetic field strength $B_{\mathrm{cr}}(a_*)$ above which no circular geodesics, timelike or null, can exist, establishing an upper magnetic bound for orbital motion. For subcritical fields, the photon circular orbit admits two real roots, the outer of which defines an outermost stable circular orbit (OSCO), complementing the conventional innermost stable circular orbit (ISCO) and confining stable motion within a finite radial domain. Exact expressions for the orbital, radial, and vertical epicyclic frequencies, and their associated precession rates, show substantial deviations from Kerr behavior, including a magnetically induced reversal of periastron precession ($\Omega_{\mathrm{per}} < 0$) within a finite radial range. 
For astrophysically relevant magnetic field strengths, the retrograde precession could be observable at large radii around astrophysical BHs, offering a potential diagnostic of large-scale magnetization.
These findings highlight the geometric influence of magnetic curvature on strong-field dynamics, providing a self-consistent framework to interpret quasi-periodic oscillation phenomenology and potential magnetic imprints in precision timing observations of compact objects.

\end{abstract}

\maketitle

\section{\label{sec:Intro} Introduction}

Magnetic fields are an integral component of the astrophysical environment surrounding black holes (BHs). Their impact extends well beyond plasma dynamics, influencing both the spacetime geometry and the motion of particles in the strong-field regime of general relativity (GR). The presence of magnetic fields in the immediate vicinity of BHs is no longer merely a theoretical possibility—it is now supported by a growing body of observational evidence. The Event Horizon Telescope (EHT), for instance, has provided compelling data through the detection of polarized emission near the supermassive BH (SMBH) M87*, constraining the magnetic field strength at event horizon scales to lie within $\sim 1–30$ G~\cite{EHT7,EHT8}. Similarly, observations of Sagittarius A* suggest that magnetic field strengths in the range of $30–100$ G are required to account for the observed synchrotron emission~\cite{eatough2013strong}. In the case of stellar-mass BHs such as Cygnus X-1, X-ray spectral modeling indicates the presence of even stronger magnetic fields, with magnitudes reaching $10^5–10^7$ G~\cite{santomnrs}. These converging lines of evidence robustly establish that BHs are generically embedded in non-negligible magnetic fields.

Magnetic fields drive a broad range of astrophysical processes: they regulate accretion dynamics as captured in magnetohydrodynamic (MHD) simulations~\cite{Gammie_2004}, shape disk behavior and synchrotron emission~\cite{Prasanna1978, Wiita1983, Iyer1985}, and shape the EHT ring morphology~\cite{EHT7,EHT8}. In curved spacetime, they also produce inherently relativistic effects, including gravitational Larmor precession \cite{CCGLP, Galtsov1978}, the gravitational Meissner effect~\cite{GME}, and the Blandford–Znajek mechanism for energy extraction~\cite{BZog}. Kerr BHs immersed in magnetic fields display additional phenomena such as corrections to gravitational Faraday rotation, the gravitational Stern–Gerlach effect~\cite{CCFaraday}, and a field-dependent efficiency of the magnetic Penrose process~\cite{CCPenrose}. Strong fields can even deform the Kerr geometry, altering photon orbits, shadow structure, and accretion features~\cite{HuoKerrMelvin, WangKerrMag}. Spin precession of test gyroscopes in magnetized Kerr spacetimes has likewise been shown to depend sensitively on the interplay between gravitoelectromagnetic (GEM) fields and external magnetism~\cite{Karthik1}.  

Recent observational progress is pushing tests of GR deeper into the strong-field regime. Measurements of orbital precession by GRAVITY~\cite{GRAVITY2020} and SINFONI~\cite{Eisenhauer2005} through S-star tracking near Sagittarius A* highlight the demand for accurate theoretical models—including those that account for external magnetic fields. Much of the literature has addressed geodesics in magnetized Schwarzschild spacetimes~\cite{Galtsov1978, Dhurandhar_1983, ZStuchlik_1999, Esteban_1984} or the dynamics of charged particles in Kerr backgrounds~\cite{bini, mbo, IyerVishveshwara} immersed in a weak magnetic field. However, the motion of neutral test particles in a generalized magnetic Kerr black hole (MKBH) remains comparatively unexplored. This case is important: it isolates the purely gravitational influence of external fields on the geometry, without the complication of Lorentz forces, and thus provides a clean testbed for linking magnetic effects to observable signatures.  

Geodesic motion encodes spacetime geometry through orbital precession. The Lense–Thirring (LT) effect, periastron precession, and radial and vertical epicyclic frequencies are central to models of quasi-periodic oscillations (QPOs), which appear as sharp features in X-ray spectra of compact objects~\cite{Stella1999, Abramowicz2001, Ingram2009}. In the relativistic precession model (RPM), low-frequency QPOs are associated with nodal precession and high-frequency QPOs with orbital and periastron motion~\cite{Stella1999, Ingram2009}. While these mechanisms have been analyzed in Kerr spacetime, the role of external magnetic fields on the fundamental frequencies of neutral geodesics has not been systematically studied. Addressing this gap is essential given the ubiquity of magnetic fields near BHs.  

In this article, we investigate the motion of neutral test particles in MKBH spacetime, focusing on how external magnetic fields modify orbital dynamics, precession, and fundamental frequencies. To this end, we adopt the Ernst~\cite{ernstBHmag} solutions—exact electrovacuum solutions to the Einstein–Maxwell equations. However, since the original Ernst spacetime exhibits well-known pathologies in its asymptotic structure, we employ the improved formulation developed in~\cite{Wild, AG2}, which remedies these issues. This improved model has already been applied to magnetic precession~\cite{AG3}, the magnetic Penrose process~\cite{CCPenrose}, and spin dynamics~\cite{Karthik1}. While astrophysical fields usually satisfy $B \ll B_{\rm max}$,
\begin{align}
B_{\rm max} \sim 2.4 \times 10^{19} \frac{M_{\odot}}{M}~{\rm G}, 
\label{bmax}
\end{align}
with $M_{\odot}$ the solar mass, studies show that for $B \sim B_{\rm max}$ the spacetime can undergo significant deformations~\cite{Galtsov1978, sha}. This motivates treating magnetic fields as an intrinsic component of the geometry rather than a perturbation in such regimes.

The structure of the paper is as follows. We begin in Sec.~\ref{sec:MagKerr} with a concise overview of the magnetized Kerr spacetime. In Sec.~\ref{sec:veffmagkerr}, we derive the effective potential governing timelike geodesics in this background. Sec.~\ref{sec:CPO} turns to the behavior of circular photon orbits and shows how the concept of a critical magnetic field naturally emerges. The idea of an outermost stable circular orbit (OSCO), distinct from the usual ISCO (innermost stable circular orbit), is introduced in Sec.~\ref{sec:OSCO}. A detailed treatment of fundamental frequencies and orbital precession appears in Sec.~\ref{sec:Fundamental Frequinces}, where we also determine the ISCO radius in the presence of a magnetic field. The magnetic field’s impact on the key orbital frequencies relevant to QPO phenomenology is systematically studied in Sec.~\ref{sec:MagQPO}. Finally, in Sec.~\ref{sec:QPOImplications}, we discuss the astrophysical implications of our framework for QPO modeling, including a justification of the spacetime for this purpose and potential observational pathways, before presenting our conclusions in Sec.~\ref{sec:dis}.

Throughout this paper, we use the metric signature $(-,+,+,+)$ and set the geometrized unit ($G = c = 1$).

\section{\label{sec:MagKerr}Kerr spacetime immersed in a uniform magnetic field}

The well-known solution describing a Kerr black hole immersed in an external uniform magnetic field provides an exact electrovacuum solution to the Einstein–Maxwell equations. This configuration preserves the axial and time-translation symmetries of the Kerr spacetime, while the magnetic field modifies its structure through the coupling between gravity and electromagnetism. The resulting metric, expressed in Boyer–Lindquist coordinates, is given as follows \cite{AG1,AG2,CCPenrose}:

\begin{align}
    ds^2 = \left( -\frac{\Delta}{A} dt^2 + \frac{dr^2}{\Delta} + d\theta^2 \right) \Sigma |\Lambda|^2 + \frac{A \sin^2\theta}{\Sigma|\Lambda|^2} (|\Lambda_0|^2d\phi - \varpi dt)^2. \label{kerrmagm}
\end{align}

The metric functions are defined as

\begin{align}
    \Delta &= r^2 + a^2 - 2Mr, & \quad \Sigma &= r^2 + a^2 \cos^2 \theta, \\
    A &= (r^2 + a^2)^2 - \Delta a^2 \sin^2 \theta, & \quad \varpi& = \frac{\alpha - \beta \Delta}{r^2 + a^2} + \frac{3}{4}a M^2 B^4. \label{varpi}
\end{align}

The functions $\alpha$ and $\beta$ in $\varpi$ are given by

\begin{align}
    \alpha &= a(1-a^2M^2B^4), \\
   \beta &= \frac{a\Sigma}{A} + \frac{aMB^4}{16} \Big(-8r\cos^2\theta (3-\cos^2\theta) - 6 r\sin^4\theta + \frac{2a^2\sin^6\theta}{A} [2Ma^2 + r(a^2+r^2)] \nonumber \\ &\quad + \frac{4Ma^2\cos^2\theta}{A} [(r^2+a^2)(3-\cos^2\theta)^2 - 4a^2\sin^2\theta]  \Big).
\end{align}

Here, $M$ represents the mass, and $a$ denotes the spin parameter of the Kerr spacetime. An asymptotically uniform magnetic field $B$ is directed along the polar axis, i.e., the vertical or
$z$-axis \cite{AG1}. The function $\Lambda(r,\theta)$ is a complex scalar that modifies the geometry due to the presence of the magnetic field. It is defined as

\begin{align}
     \Lambda  &\equiv \Lambda (r,\theta) = \text{Re} \ \Lambda + i \ \text{Im} \ \Lambda \nonumber \\ 
     & = 1 + \frac{B^2 \sin^2\theta}{4} \left [(r^2 +a^2) + \frac{2a^2Mr\sin^2\theta}{\Sigma} \right ] - i \frac{aB^2 M \cos\theta}{2} \left (3 - \cos^2\theta + \frac{a^2 \sin^4\theta}{\Sigma} \right ).
\end{align}

An essential feature of this metric is the Harrison–Ernst function $\Lambda(r, \theta)$, which appears before $d\phi$ in Eq.~\eqref{kerrmagm}. This function ensures the metric remains a valid solution to the Einstein–Maxwell equations in the presence of a magnetic field. Specifically, the factor $|\Lambda_0|^2$ is introduced to regularize the geometry along the polar axis ($\theta \rightarrow 0$), thereby eliminating conical singularities \cite{William1981, AG1, AG2}. This correction prevents divergences in the Ricci tensor and guarantees the physical consistency of the spacetime. To maintain regularity at the axis, the azimuthal coordinate range must also be rescaled from $2\pi$ to $2\pi |\Lambda_0|^2$ \cite{AG1, AG2}, where
\begin{align}
    |\Lambda_0|^2 = |\Lambda(r,0)|^2 = 1 + a^2 M^2 B^4.
\end{align}
In the absence of this rescaling, the metric would no longer fulfill the Einstein–Maxwell equations because conical singularities would emerge at the poles. As mentioned above, due to the regularization of the symmetry axis, the azimuthal coordinate is rescaled such that 
\(\phi \in [0, 2\pi |\Lambda_0|^2]\). 
Consequently, the physically observed azimuthal frequency ($\Omega_{\phi}^{\rm obs}$) differs from the coordinate frequency ($\Omega_{\phi}\equiv d\phi/dt$) by a constant factor, i.e., 
$\Omega_{\phi}^{\rm obs} = \Omega_{\phi}/|\Lambda_0|^2$. In the weak magnetic field ($B \ll M^{-1}$) and/or slow-rotation ($a/M \ll 1$) limits, one has 
$|\Lambda_0|^2 \to 1$, and the distinction between coordinate and observed frequencies becomes negligible.

Notably, the structure of the event horizons remains unaffected by the presence of the external magnetic field. The radial locations of the outer and inner horizons are still determined by the roots of $\Delta = 0$:
\begin{align}
    r_\pm = M \left(1 \pm \sqrt{1-a_*^2} \right),
\end{align}
where $a_*=a/M$. Thus, the positions of the horizons are independent of the magnetic field strength.

However, the ergoregion of the MKBH is influenced by the magnetic field. The boundary of the ergoregion is defined by the surface where $g_{tt} = 0$, leading to the condition
\begin{align}
    g_{tt} = - \frac{\Delta \Sigma}{A} |\Lambda|^2 + \frac{\varpi^2 A \sin^2\theta}{\Sigma |\Lambda|^2} = 0.
    \label{MKBHergo}
\end{align}
This equation cannot be solved analytically but can be investigated numerically to reveal the deformation of the ergosurface as a function of the magnetic field strength \cite{Gibbons2013}.

Finally, we emphasize that the magnetized Kerr spacetime is not asymptotically flat. As $r \to \infty$, the magnetic field does not vanish; instead, it gradually dominates the curvature of the spacetime. In this limit, the geometry approaches that of the Melvin universe \cite{melvinverse, Melvin1}, an exact, static, and axisymmetric solution of the Einstein–Maxwell equations representing a spacetime filled with a uniform magnetic field. However, as we justify in Sec. \ref{sec:QPOImplications}, it serves as an excellent model for the strong-field region near the BH where phenomena like QPOs originate.

\section{\label{sec:veffmagkerr}Effective Potential in MKBH Spacetime}

In this section, we analyze the motion of neutral test particles confined to the equatorial plane of the MKBH spacetime. While the geodesic structure of magnetized Schwarzschild spacetimes has been extensively explored in the literature~\cite{Galtsov1978, Dhurandhar_1983, ZStuchlik_1999, Esteban_1984}, the rotating case introduces significant complexities. Although the spacetime retains integrability via conserved quantities associated with its symmetries, the Hamilton–Jacobi equation becomes non-separable, thereby complicating the extraction of exact analytic solutions. Nevertheless, for our purposes—restricted to equatorial dynamics—a tractable and self-consistent framework remains available.

We begin with a general stationary, axisymmetric spacetime described by the line element

\begin{align}
ds^2 &= g_{tt} dt^2 + 2g_{t\phi} dt\, d\phi + g_{\phi\phi} d\phi^2 + g_{rr} dr^2 + g_{\theta\theta} d\theta^2\,.
\end{align}

The motion of a test particle along a geodesic, with four-velocity $u^\mu = \dot{x}^{\mu} = dx^{\mu}/d\lambda$, obeys the geodesic equation $u^\mu \nabla_\mu u^\nu = 0$. This motion can be derived from the Lagrangian~\cite{StraumannGR}

\begin{align}
    \mathcal{L}(x^\mu, \dot{x}^\mu) = \frac{1}{2} g_{\mu \nu} \dot{x}^\mu \dot{x}^\nu,
\end{align}

which leads to the Hamiltonian

\begin{align}
    \mathcal{H}(x^\mu, p_\mu) = p_\mu \dot{x}^\mu - \mathcal{L}(x^\mu, \dot{x}^\mu) = \frac{1}{2} g^{\mu \nu} p_\mu p_\nu,
    \label{eq:H1}
\end{align}

where the conjugate momenta are defined as $p_\mu = \partial \mathcal{L}/\partial\dot{x}^\mu = g_{\mu \nu} \dot{x}^\nu$. The dynamics are then governed by the Hamiltonian equations~\cite{StraumannGR},

\begin{align}
    \dot{x}^\mu = \frac{\partial \mathcal{H}}{\partial p_\mu}, \quad \dot{p}_\mu = - \frac{\partial \mathcal{H}}{\partial x^\mu}.
\end{align}

Stationarity and axisymmetry imply the existence of two Killing vector fields, $\partial_t$ and $\partial_\phi$, yielding two conserved quantities:

\begin{align}
E &= -p_t = -g_{tt} \dot{t} - g_{t\phi} \dot{\phi} \label{eq:metricEL1}\,, \\
L &= p_\phi = g_{t\phi} \dot{t} + g_{\phi\phi} \dot{\phi} \label{eq:metricEL2}\,.
\end{align}

These are typically interpreted as the conserved energy and angular momentum per unit mass of the particle. However, such an interpretation strictly assumes asymptotic flatness, which is not valid for the non-asymptotically flat MKBH geometry. Consequently, the physical meaning of $E$ and $L$ becomes more subtle, as there is no preferred reference frame at spatial infinity.

Solving Eqs.~\eqref{eq:metricEL1}–\eqref{eq:metricEL2} for $\dot{t}$ and $\dot{\phi}$ gives

\begin{align}
\dot{t} = \frac{E g_{\phi\phi} + L g_{t\phi}}{d}, \quad 
\dot{\phi} &= \frac{-E g_{t\phi} - L g_{tt}}{d} ,
\end{align}

where we define

\begin{align}
    d \equiv g_{t\phi}^2 - g_{tt} g_{\phi\phi}.
\end{align}

Restricting to equatorial motion ($\theta = \pi/2$) with $\dot{\theta} = 0$, we use the Hamiltonian constraint from Eq.~\eqref{eq:H1} to obtain

\begin{align}
    2\mathcal{H} = g^{rr} p^2_r - \frac{g_{\phi \phi} E^2 + 2 g_{t \phi} E L + g_{tt} L^2}{d} = \kappa,
\end{align}

where $\kappa=-1$ for timelike and $\kappa \rightarrow 0$ for null geodesics. Substituting $p_r = g_{rr} \dot{r}$ and simplifying yields

\begin{align}
\dot{r}^2 &= \frac{1}{g_{rr} \cdot d} \left[ g_{\phi \phi} E^2 + 2 g_{t\phi} EL+g_{tt} L^2  \right] + \frac{\kappa}{g_{rr}}.
\label{eq:rdot2}
\end{align}

After some algebra, Eq.~\eqref{eq:rdot2} can be recast in the useful form

\begin{align}
    V_\mathrm{r} (r) = \dot{r}^2 = \frac{g_{\phi \phi}}{g_{rr} \cdot d} (E-V_+) (E-V_-) + \frac{\kappa}{g_{rr}},
    \label{eq:Veff1}
\end{align}

where \cite{Taylor2025B}

\begin{align}
    V_\pm = \frac{L}{g_{\phi\phi}} \Big\{ - g_{t\phi} \pm \sqrt{d} \Big\}. \label{eq:V+-}
\end{align}

Although $V_\pm$ is often referred to as an ``effective potential” in the literature \cite{PuglieseK, PuglieseKN}, we reserve the term effective potential for $V_\text{eff}$. To distinguish between them, we follow the terminology of Ref.\cite{Frolov2011} and refer to $V_\pm$ as potential functions.
For a detailed discussion of the qualitative features of these functions and their physical interpretation, we refer the reader to Refs.\cite{Frolov1998, Frolov2011, ferrari2020GR} and references therein. 

We now specialize to the motion of a neutral particle along a timelike geodesic. Setting $\kappa = -1$ in Eq.~\eqref{eq:Veff1}, and applying this to the equatorial MKBH spacetime, we obtain

\begin{align}
    V_\mathrm{r} (r) = \frac{A_e}{\Lambda_e^4 r^4} (E-V_-)(E-V_+) - \frac{\Delta}{\Lambda_e^2 r^2},
    \label{eq:VeffMKBH}
\end{align}
where, $A_e$ and $\Lambda_e$ represent the functions $A$ and $\Lambda$ evaluated at the equatorial plane ($\theta = \pi/2$): 
\begin{align*}
A_e = (r^2+a^2)^2 - \Delta a^2, \quad \Lambda_e = 1 + \frac{B^2}{4} \left( r^2 + a^2 + \frac{2a^2M}{r} \right),
\end{align*}
consistent with the definitions provided in Sec.~\ref{sec:MagKerr}.
In the appropriate limits—namely Kerr, magnetized Schwarzschild, and the Melvin universe—Eq. (\ref{eq:VeffMKBH}) reduces to known results, as shown in Appendix \ref{app:Vefflimits}, thereby validating the consistency of our formulation.

\begin{figure}[H]
\begin{center}
\subfigure[\ $a_*=0.1$]{
\includegraphics[width=2.7in,angle=0]{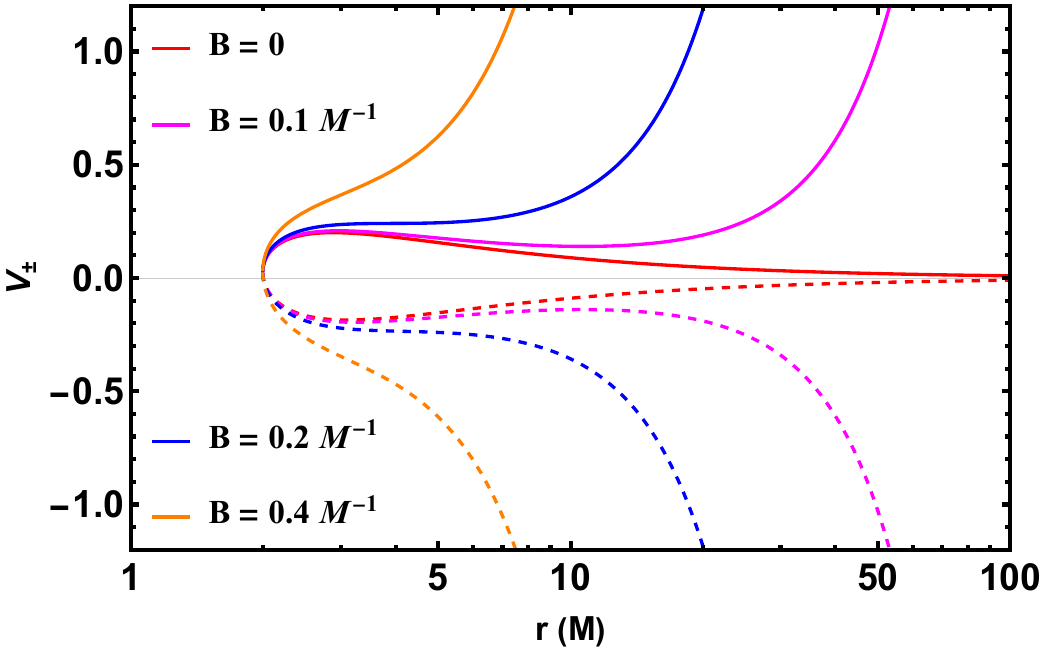}}
\subfigure[\ $a_*=0.9$]{
\includegraphics[width=2.7in,angle=0]{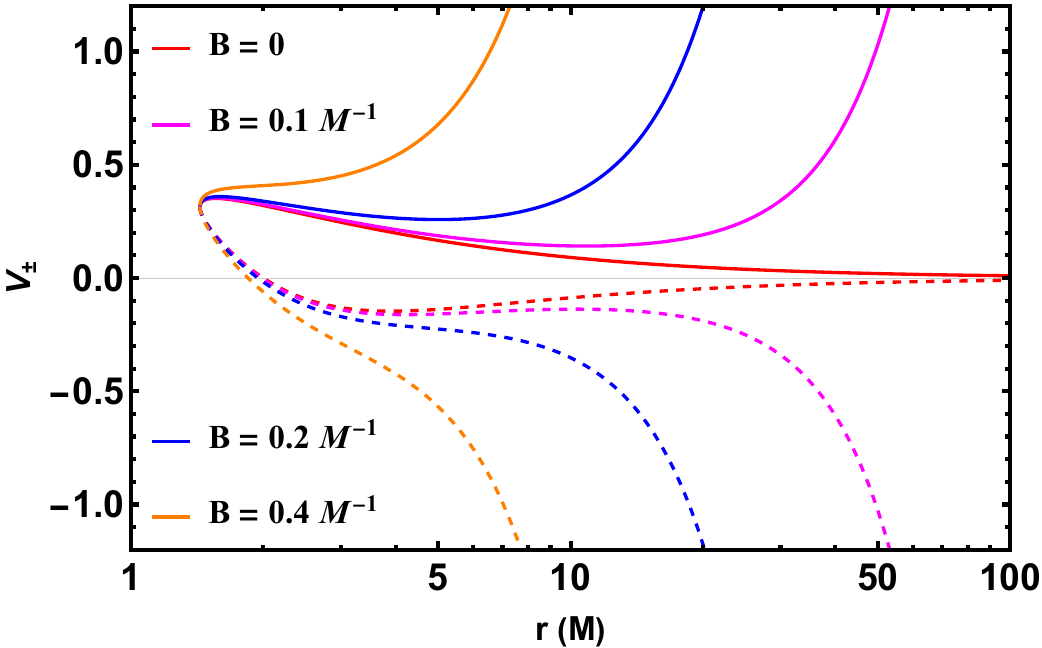}}
\caption{\label{fig:PFa} The potential functions $V_+(r)$ (solid curves) and $V_-(r)$ (dashed curves) for prograde equatorial orbits in the MKBH spacetime, with spin parameters (a) $a_* = 0.1$ and (b) $a_* = 0.9$ for different magnetic field strengths $B$ are plotted. See Sec.~\ref{sec:veffmagkerr} for details.}
\end{center}
\end{figure}

Figure~\ref{fig:PFa} illustrates the characteristic behavior of the potential functions \( V_{\pm}(r) \) for various magnetic field strengths, highlighting the phenomenon of magnetic confinement. In the pure Kerr case (\( B = 0 \)), the potential functions approach finite asymptotic values, as expected for an asymptotically flat spacetime. However, with increasing \( B \), the potential grows without bound at large radii, forming a pronounced barrier. At sufficiently high magnetic field strengths, this barrier becomes so strong that it prevents even photons from crossing it. A similar effect was reported earlier for the magnetized Schwarzschild spacetime~\cite{Galtsov1978, BHshield}.  

This behavior indicates the presence of a critical magnetic field strength, \( B_{\text{cr}} \), beyond which no circular orbits can exist. In essence, the magnetic field acts as a confining ``wall,'' restricting particles to a finite radial domain and drastically modifying the orbital dynamics compared to the unmagnetized case.  

A closer inspection of panels (a) and (b) in Fig.~\ref{fig:PFa} shows that the influence of the spin parameter \( a_* \) on the potential barrier is relatively minor. The \( V_{\pm} \) profiles remain symmetric when \( a_* \to 0 \), but this symmetry is broken for nonzero spin. The degree of asymmetry increases progressively with larger values of \( a_* \).  

We now proceed to discuss the determination and physical implications of \( B_{\text{cr}} \) in detail in the next section.

\section{\label{sec:CPO} Circular Photon Orbit and critical magnetic field}

The presence of a circular photon orbit (CPO) encodes essential properties of the spacetime geometry, serving as a key indicator in distinguishing compact objects. It significantly influences observational signatures, including quasi-normal mode spectra and BH shadow profiles. Their location therefore provides a useful probe of both theoretical models and astrophysical observations.

To determine the radius of the CPO in a magnetized Kerr geometry, we follow the method outlined by Bardeen~\cite{Bardeen1972,ShapTeuBook}.The conserved quantities $E$ [Eq.\eqref{eq:Emetric}] and $L$ [Eq.\eqref{eq:Lmetric}] imply that the denominators must remain real for physically meaningful trajectories. The limiting case, where the denominator vanishes, defines the CPO radius, yielding

\begin{align}
    & r^4 \Big(M (5 B^2 r^2-12)-3 B^2 r^3+4 r \Big) 
    + a^6 B^6 M^3 (2 M+r) (M+5 r) \nonumber 
    \\ & 
    + a^4 B^2 M \Big(M r (11-3 B^4 r^4)+B^2 M^3 r (11 B^2 r^2-12)
    +2 M^2 (2 B^4 r^4-6 B^2 r^2+1)+5 r^2\Big) \nonumber 
    \\ & 
    + a^2 r \Big(4 M r (B^2 r^2-3)-3 B^2 r^4+B^4 M^3 r^3 (5 B^2 r^2-12)
    +M^2 (-3 B^6 r^6+4 B^4 r^4+11 B^2 r^2-12)\Big) \nonumber
    \\ &
    + 2 (1+a^2 B^4 M^2) (a^3 M + 3 a M r^2) 
    \sqrt{\Big(B^2 (a^2 (2 M+r)+r^3)-4 r\Big) \Big(M (3 B^2 (a^2+r^2)-4)-2 B^2 r^3\Big)} = 0.
    \label{cpomagkerr}
\end{align}

In the Schwarzschild limit ($a \to 0$), this reduces to

\begin{align}
3  B^2  r^3 - 5 M B^2 r^2 -  4 r + 12 M = 0,
\label{CPOEMS BH}
\end{align}

which reproduces the CPO equation for the magnetized Schwarzschild BH (EMS BH)\cite{Galtsov1978}. Eq.\eqref{CPOEMS BH} admits two positive real roots for $0 < B < B_{\text{cr}}$, a single root at $B=B_{\text{cr}}$, and none for $B > B_{\text{cr}}$, establishing a critical field $B_{\text{cr}} = 0.189 M^{-1}$ above which no circular orbits exist \cite{Galtsov1978,BHshield}. 

\begin{figure}[h!]
\begin{center}{
\includegraphics[width=4in,angle=0]{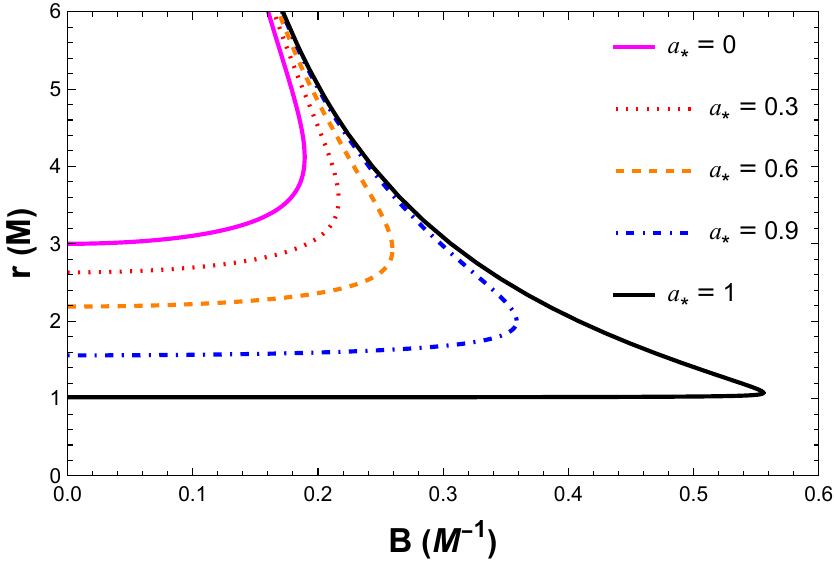}}
\caption{\label{fig:Bcric} CPO radius $r (M)$ as a function of $B (M^{-1})$ for various spin parameter $a_*$. Turning point ($\equiv B_\text{cr}$) of each curve marks the critical field strength beyond which no circular orbits exist. See Sec.\ref{sec:CPO} for further details.}
\end{center}
\end{figure}

\begin{figure}[!h]
\begin{center}{
\includegraphics[width=4in,angle=0]{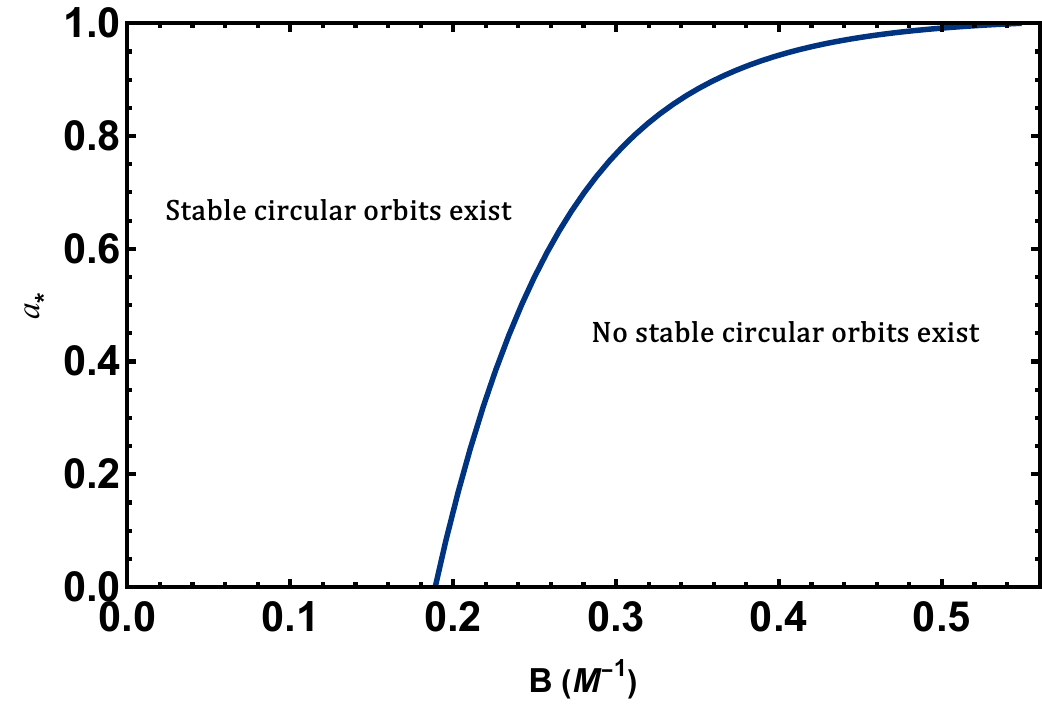}}
\caption{\label{fig:Bvsa} Parameter space for stable circular orbits in the equatorial plane. The solid curve divides the $B-a_*$ plane into regions where circular orbits are permitted (left) and forbidden (right). The endpoints of the curve are located at $(a_*,B) \sim (0, 0.189 M^{-1})$ and $(1, 0.556 M^{-1})$. See Sec.\ref{sec:CPO} for further details.}
\end{center}
\end{figure}

For the MKBH, a similar threshold exists, though closed-form solutions are not tractable. We therefore solve Eq.~\eqref{cpomagkerr} numerically to find the physically relevant roots. Our analysis, summarized in Fig.~\ref{fig:Bcric} and Table-\ref{tab:B_critical}, identifies a turning point on each curve that defines the critical magnetic field strength $B_{\text{cr}}$ for a given spin; this critical value increases with $a_*$. We map this entire relationship across the parameter space in Fig.~\ref{fig:Bvsa}. The solid curve of $B_{\text{cr}}(a_*)$ divides the $B-a_*$ plane, implying that sufficiently strong magnetic fields ($B > B_{\text{cr}}$), corresponding to the region right of the curve, prohibit all circular orbits, thereby halting any associated physical processes \cite{WangKerrMag,Zhang2024}. The physical mechanism for this prohibition could be demonstrated using Eq.\eqref{eq:V+-} by substituting the metric components in the same. As shown in Fig.~\ref{fig:PFa}, the potential function becomes arbitrarily large if $B$ is in the order of $B_\text{cr}$ and closed to it, preventing any test particles (even photons) from overcoming the potential barrier. For detail, one may also look into \cite{BHshield} for the physical meaning of large potential function, which was described for the magnetic Schwarzschild BH. However, from the theoretical perspective, this sets an upper bound on the allowed magnetic field strength for a given $a_*$.

\begin{table}[h!]
\centering
\setlength{\tabcolsep}{8pt} 
\renewcommand{\arraystretch}{1.3}
\begin{tabular}{|c|c|c|c|c|c|c|c|c|c|c|c|}
\hline
$a_*$ & 0.0 & 0.1 & 0.2 & 0.3 & 0.4 & 0.5 & 0.6 & 0.7 & 0.8 & 0.9 & 1.0 \\
\hline
$B_\text{cr} \ (M^{-1})$ &
0.189 & 0.197 & 0.206 & 0.216 & 0.228 & 0.242 & 0.259 & 0.281 & 0.311 & 0.359 & 0.556 \\
\hline
\end{tabular}
\caption{Critical magnetic field $B_{\text{cr}}$ for MKBH spacetime. See Sec.~\ref{sec:CPO} for details.}
\label{tab:B_critical}
\end{table}

This structure of the CPO shows that circular geodesics in the magnetized Kerr background are not unbounded, unlike in the standard Kerr case. Instead, they are confined within a finite radial interval, pointing to the existence of an outermost limit to admissible circular motion. The nature of this outer boundary is clarified in the next section.

\section{\label{sec:OSCO} Outermost Stable Circular Orbits}

In asymptotically flat spacetimes, such as the standard Kerr geometry, stable circular orbits exist from the ISCO out to infinity. However, as noted in Sec.~\ref{sec:MagKerr}, the magnetized Kerr spacetime is non-asymptotically flat and approaches the Melvin universe as ($r\to\infty$). This change in asymptotics introduces a novel feature absent in the standard Kerr case: an \textit{outermost stable circular orbit} (OSCO) located far from the BH horizon. The OSCO is distinct from the ISCO: while inside the ISCO unstable circular orbits may exist, beyond the OSCO no circular motion at all (timelike or null) is allowed. In fact, the existence of an ISCO in magnetized black hole spacetimes is necessarily accompanied by an OSCO \cite{HuoKerrMelvin}, because the magnetic confinement of the Melvin asymptotics restricts admissible trajectories at large radii.

To build intuition, consider the EMS BH limit where the CPO equation, Eq.~\eqref{CPOEMS BH}, admits two positive real roots for $B < B_{\text{cr}}$ \cite{Galtsov1978}. These roots are approximately
\begin{align}
r_1 \approx 3 M + 9M \left(\frac{B}{B_{\rm max}} \right)^2, \quad r_2 \approx \frac{2}{\sqrt{3}} \frac{1}{B},
\label{eq:r1r2}
\end{align}
and, we identify the larger root $r_2$ with the OSCO, i.e., $r_{\text{OSCO}} \equiv r_2$. In the limit $B \to 0$ one recovers the Schwarzschild results $r_1 \to 3M$ and $r_{\text{OSCO}} \to \infty$. Similar results were obtained in Refs.~\cite{Esteban_1984,Dadhich1979,Galtsov1978,AlievMagBHrev,BHshield}.

\begin{figure}[H]
\begin{center}
\subfigure[$a_*=0$]{
\includegraphics[width=2.7in,angle=0]{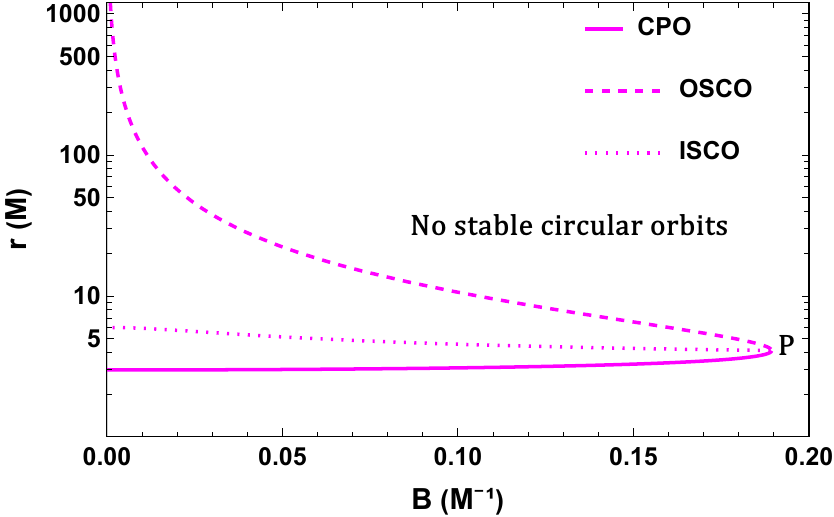}}
\subfigure[$a_*=0.3$]{
\includegraphics[width=2.7in,angle=0]{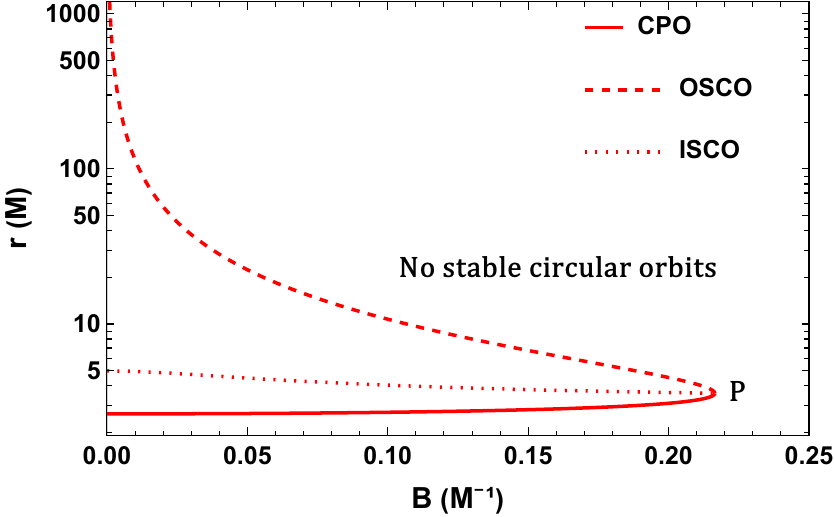}}
\subfigure[$a_*=0.6$]{
\includegraphics[width=2.7in,angle=0]{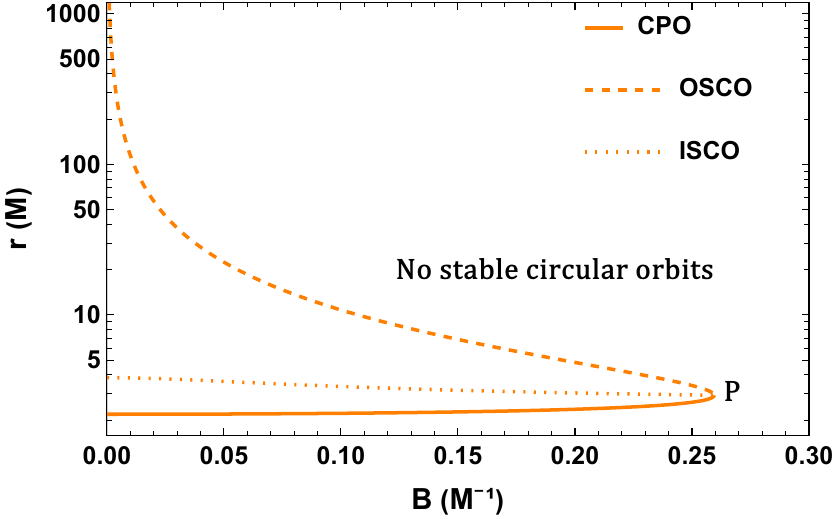}}
\subfigure[$a_*=0.9$]{
\includegraphics[width=2.7in,angle=0]{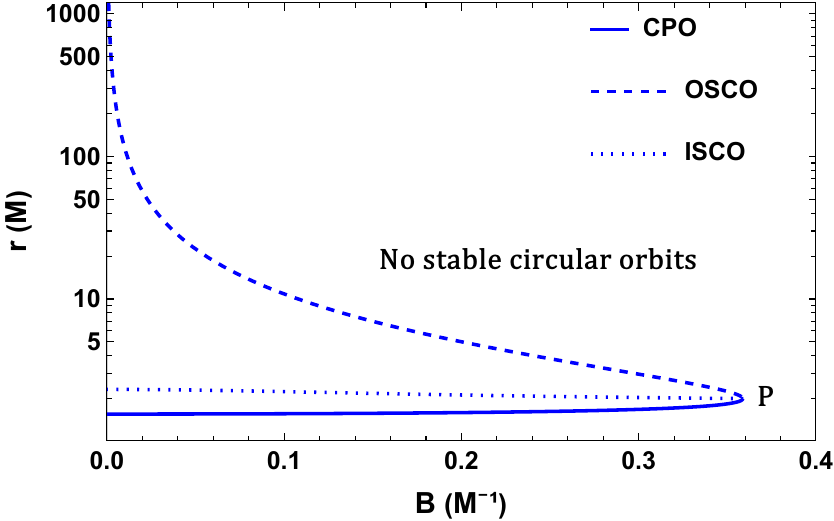}}
\subfigure[$a_*=1$]{
\includegraphics[width=2.7in,angle=0]{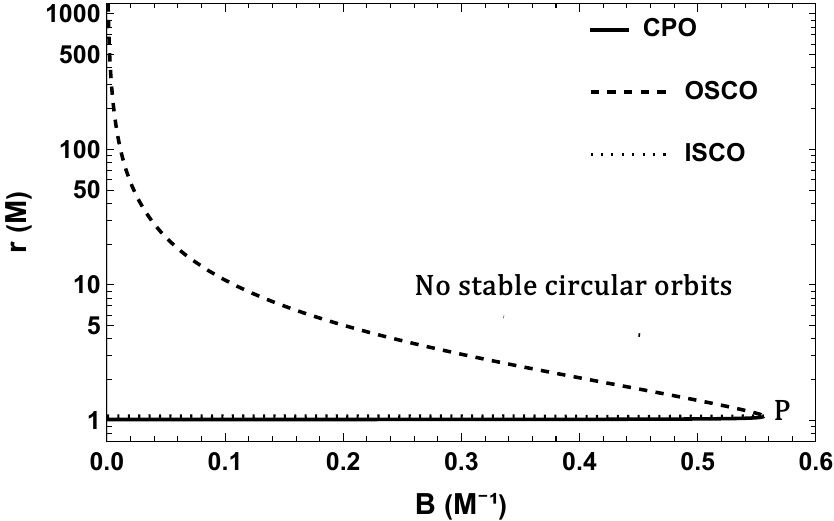}}
\caption{\label{Fig:CPO} $r$ (in $M$) as a function of $B$ (in $M^{-1}$) for MKBH with different spin parameters $a_*$. For each value of $a_*$, the CPO equation yields two branches, with the outer one identified as OSCO. We have drawn ISCO (see Sec. \ref{subsec:ISCO}) for reference. At the critical field $B_{\text{cr}}$ all three coincide at point P. Beyond this point, no circular orbits—timelike or null—are allowed. See Sec.\ref{sec:OSCO} and Sec. \ref{subsec:ISCO} for details.}
\end{center}
\end{figure}

For the MKBH geometry, simple closed-form  analytical solutions are not possible, but the qualitative picture remains: the CPO equation yields two positive branches for $B < B_{\rm cr}$. Both correspond to circular photon orbits, but the outer branch is identified as the OSCO. We propose a physically motivated rationale for this identification: if null geodesics—the trajectories of photons—are not permitted in circular configurations beyond this outer CPO root, then no stable circular timelike geodesic can exist there either. 
It is important to make explicit the asymmetry between null and timelike confinement: the domain of stable timelike circular motion is necessarily more restricted, and is bounded above by the outer root of CPO. This follows because timelike circular orbits must additionally satisfy the radial stability condition, $\Omega_r^2 \ge 0$, whereas null circular orbits correspond to the limiting case without this requirement. This provides a clear physical basis for interpreting the outer CPO root as a definitive outer boundary of stable motion, thereby confining astrophysical phenomena such as accretion disks and binary orbits to a finite region between the ISCO and the OSCO. The location of the OSCO depends sensitively on both $B$ and $a_*$: lower $B$ field strengths push the OSCO outward, while the stronger $B$ fields pull it inward, reflecting the fact that the magnetic field effectively confines the circular motion at large radii.  At the critical field $B = B_{\rm cr}$, the CPO, the ISCO, and the OSCO coincide at a single radius. We denote this coincidence by point $P$ in Fig.~\ref{Fig:CPO}; at $P$, the three characteristic radii merge, marking the disappearance of any radius that can support stable timelike circular motion for $B > B_{\rm cr}$. Beyond this point, no circular orbits—timelike or null—exist (see Fig.~\ref{Fig:CPO} and the labeled point $P$). 
Our numerical evaluation of the marginal stability condition, $\Omega_r^2 = 0$, confirms that this bound is saturated at point $P$, where the ISCO, OSCO, and CPO coincide, marking the disappearance of the stable orbital window. Consequently, the OSCO complements the ISCO in delimiting the physically accessible region for circular geodesics in magnetized Kerr spacetime. All further analysis of orbital frequencies and precession are therefore be restricted to the physically admissible window:
$r_{\rm ISCO} \le r < r_{\rm OSCO}$. We now proceed to investigate how the ISCO radius varies with $a$ and $B$.

\section{\label{sec:Fundamental Frequinces}Fundamental Frequencies and Orbital Precession}

The motion of a test particle is characterized by three fundamental frequencies, which play a central role in various astrophysical scenarios such as accretion disk dynamics and QPOs. These include the Keplerian frequency $\Omega_\phi$, the vertical epicyclic frequency $\Omega_\theta$, and the radial epicyclic frequency $\Omega_r$. While the Keplerian frequency governs the azimuthal motion, the epicyclic frequencies arise from small oscillations about a stable circular orbit and are intimately connected to precession effects. The latter two frequencies are related to the precession of the orbit and orbital plane. The periastron precession frequency $\Omega_\text{per}$ describes the precession of the orbit, whereas the nodal plane precession frequency $\Omega_\text{nod}$ corresponds to the precession of the orbital plane itself. In this section, we derive expressions for these fundamental frequencies and determine the ISCO radius in the MKBH spacetime. In what follows, the orbital and epicyclic frequencies are defined with respect to the coordinate time $t$ associated with the stationary Killing vector $\partial_t$, corresponding to a distant Killing-time normalization. This provides a consistent global reference for defining frequencies even in the absence of asymptotic flatness. 

\subsection{\label{sec:analyticfreqs}Analytical expressions for fundamental orbital frequencies}

To evaluate the fundamental frequencies in the magnetized Kerr spacetime, we substitute the corresponding metric components $g_{\mu\nu}$ into the general expressions mentioned in Appendix~\ref{app:general}. The orbital
frequency for the prograde orbit can be written as \footnote{We present the analytical expressions up to $\mathcal{O}(B^4)$ to illustrate the dependence of observables on the magnetic field. The full exact expressions, used to generate all numerical results and plots, are provided in a publicly available supplementary Mathematica notebook accompanying this work.}

\begin{align}
\Omega_\phi \approx \Omega^{\text{Kerr}}_\phi + \frac{\left(2 a^2 M + a (r - 5M) \sqrt{M r} + r^2 (M + r)\right)}{4 \left(a M + \sqrt{M} r^{3/2}\right)} B^2 + \mathcal{O}(B^4), \label{magekerrkep}
\end{align}
where \cite{CCDKNS}
\begin{align}
\Omega^{\text{Kerr}}_\phi = \frac{M^{1/2}}{a M^{1/2} + r^{3/2}}.
\end{align}

Similarly, the radial and vertical epicyclic frequencies are expressed as
\begin{align}
    \Omega_r &\approx  \Omega^{\text{Kerr}}_r + \bigg(2 a^4 M (4 M+3 r)+a^3 \left(-20 M^2-21 M r+3 r^2\right) \sqrt{M r} + 2 a^2 M r \left(6 M^2+16 M r-r^2\right) \nonumber \\ & +a \sqrt{M} r^{5/2} \left(-18 M^2-19 M r+11 r^2\right)+4 r^4 \left(6 M^2-5 M r+r^2\right)\bigg) B^2 \Bigg/ \left(4 r\left(a \sqrt{M} r+r^{5/2}\right)^2 \Omega^{\text{Kerr}}_r\right) \nonumber \\ &+ \mathcal{O}(B^4), \label{magekerrR}
\end{align}

where,  \cite{CCDKNS}

\begin{align}
     \Omega^{\text{Kerr}}_r &= \frac{ \Omega^{\text{Kerr}}_\phi}{r} \left(r^2 - 6 Mr+8ar^{1/2}M^{1/2}-3a^2 \right)^{1/2},
\end{align}

and,

\begin{align}
    \Omega_\theta \approx  \Omega^{\text{Kerr}}_\theta - \frac{a \sqrt{M} \left(2 a \sqrt{M}+\sqrt{r} (r-3 M)\right) \left(a^2 (2 M+3 r)+3 r^3\right)}{4 r \left(a \sqrt{M} r+r^{5/2}\right)^2 \Omega^{\text{Kerr}}_\theta} B^2 + \mathcal{O}(B^4), \label{magekerrV}
\end{align}
where,  \cite{CCDKNS}
\begin{align}
    \Omega^{\text{Kerr}}_\theta &= \frac{ \Omega^{\text{Kerr}}_\phi}{r} \left(r^2 - 4aM^{1/2}r^{1/2} + 3a^2 \right)^{1/2}.
\end{align}
respectively. Likewise, using Eqs.~\eqref{Appendixper} and \eqref{Appendixnod}, we obtain the analytical expressions for the periastron and nodal precession frequencies in the magnetized Kerr spacetime. Expanding up to $\mathcal{O}(B^2)$, the periastron precession frequency takes the form
\begin{align}
    \Omega_\text{per} &\approx \Omega_{\text{per}}^{\text{Kerr}} + \frac{1}{4} B^2 \frac{N_1}{(aM+\sqrt{M}r^{3/2})} + \mathcal{O}(B^4), \label{magekerrper}
\end{align}

where, $\Omega_{\text{per}}^{\text{Kerr}} = \Omega_\phi^{\text{Kerr}} - \Omega_r^{\text{Kerr}}$, and the magnetic correction term $N_1$ is given by

\begin{eqnarray} \nonumber
    N_1 &=& \left[2 a^2 M+a (r-5 M) \sqrt{M r}+r^2 (M+r) \right]-\frac{1}{r^2 \sqrt{-3 a^2+8 a \sqrt{M r}+r (r-6 M)}} .
    \\
    && \bigg[2 a^4 M (4 M+3 r)+a^3 \left(-20 M^2-21 M r+3 r^2\right) \sqrt{M r}  +2 a^2 M r \left(6 M^2+16 M r-r^2\right)\nonumber
    \\
    &+& a \sqrt{M} r^{5/2} \left(-18 M^2-19 M r+11 r^2\right)+4 r^4 \left(6 M^2-5 M r+r^2\right)\bigg].
\end{eqnarray}
Similarly, the nodal plane precession frequency is given by

\begin{align}
    \Omega_{\text{nod}} \approx \Omega_{\text{nod}}^{\text{Kerr}} + \frac{1}{4} B^2 \frac{N_2}{(aM+\sqrt{M}r^{3/2})} + \mathcal{O}(B^4), \label{magekerrnod}
\end{align}

where, $\Omega_{\text{nod}}^{\text{Kerr}} = \Omega_\phi^{\text{Kerr}} - \Omega_\theta^{\text{Kerr}}$, and the magnetic correction term $N_2$ reads

\begin{align}
    N_2 = \left[2 a^2 M+a (r-5 M) \sqrt{M r}+r^2 (M+r) \right]  + \frac{a \left[2 a M+\sqrt{Mr} (r-3 M)\right] \left[a^2 (2 M+3 r)+3 r^3\right]}{r^2 \sqrt{3 a^2-4 a \sqrt{M r}+r^2}}.
\end{align}

These precession frequencies are of particular astrophysical relevance, especially in modeling QPOs and disk dynamics. To build physical insight, we next explore how these frequencies behave in various limiting cases — such as the magnetized Schwarzschild and Melvin spacetime — before turning to a detailed analysis of magnetic field effects in Sec.~\ref{sec:MagQPO}.

\subsection{\label{subsec:ISCO}Innermost Stable Circular Orbit in Magnetized Kerr Spacetime}

The innermost stable circular orbit (ISCO), denoted by $r_{\text{ISCO}}$, marks the smallest radial distance at which a test particle can stably orbit a BH. For $r < r_{\text{ISCO}}$, circular orbits become unstable, causing particles to plunge into the BH. In MKBH spacetime, determining $r_{\text{ISCO}}$ analytically is non-trivial due to the coupling between the BH's spin $a$ and the external magnetic field $B$. Following the approach of \cite{OrbitalDoneva}, we compute $r_{\text{ISCO}}$ numerically by solving the condition for radial stability $\Omega_r^2=0$. This condition ensures that infinitesimal radial perturbations do not grow, preserving orbital stability.  

\begin{figure}[H]
    \centering
    \includegraphics[width=0.7\linewidth]{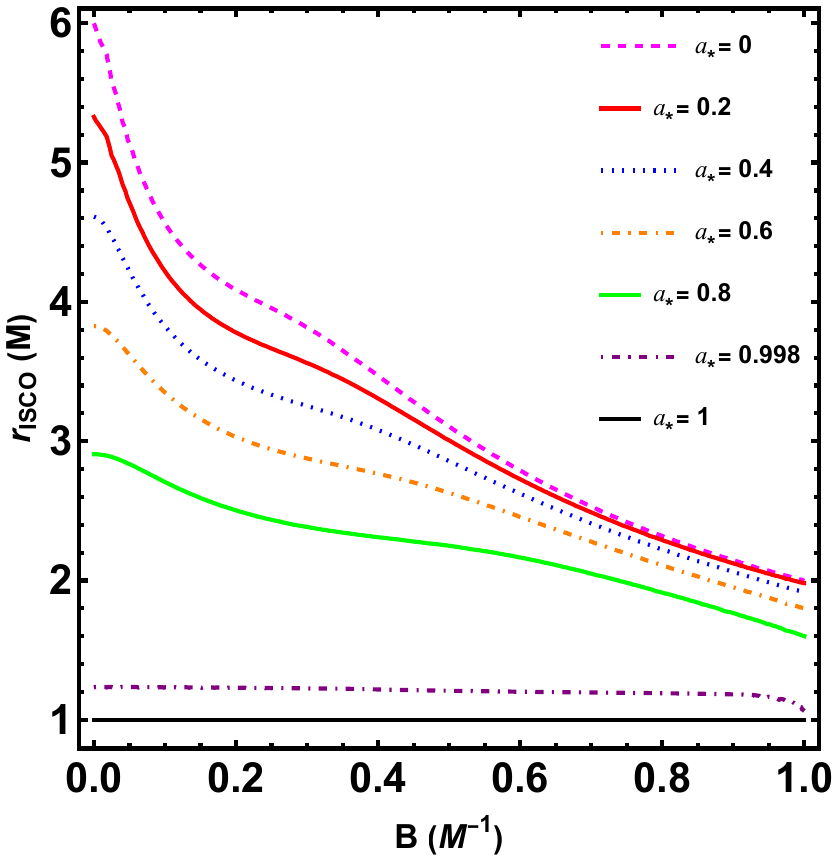}
    \caption{\label{fig:ISCO} $r_{\text{ISCO}}$ (in units of $M$) as a function of $B$ (in $M^{-1}$) for varying spin parameter $a_*$. The ISCO radius reaches a maximum at $B = 0$ and decreases monotonically with increasing $B$, except for the extremal case. See Sec.\ref{subsec:ISCO} for further details}
\end{figure}

As shown in Fig.~\ref{fig:ISCO}, the ISCO radius $(r_{\text{ISCO}})$ decreases with increasing magnetic field strength for all sub-extremal spin values $a_* < 1$, reflecting the influence of the magnetic field. However, the extremal case ($a_* = 1$) exhibits markedly different behavior, where $r_{\text{ISCO}}$ remains fixed at the horizon radius $R_h = M$ for all values of $B$, showing no dependence on the magnetic field. This is because the magnetic field is effectively expelled from the vicinity of an extremal Kerr BH — a manifestation of the well-known \textit{gravitational Meissner effect} \cite{WaldBHmag, King1975, Bicak1985}.

\subsection{\label{Sec:Fundamental Special} Special Cases}

\subsubsection{\textbf{Magnetized Schwarzschild spacetime ($a \rightarrow 0,B\neq0$)}}

Now, in the magnetized Schwarzschild spacetime $(a \rightarrow 0)$ — also known as the Ernst-Melvin-Schwarzschild (EMS) spacetime \cite{ernstBHmag} Fundamental precession frequencies are given by \cite{CCGLP}

\begin{align}
    \Omega_\phi^{\text{EMS}} &= \frac{(4+B^2 r^2)^2}{16} \sqrt{\frac{M(4-3B^2r^2)+2B^2r^3}{r^3(4-B^2r^2)}} \\
\Omega_r^{\text{EMS}} &=  \frac{1}{r^2 (4 + B^2 r^2)}\sqrt{\frac{N}{(4 - B^2 r^2)}}, \\ 
\Omega_{\theta}^{\text{EMS}} &= \sqrt{\frac{M}{r^3}},
\end{align}

where

\begin{align}
    N= & {} 4 B^2 r^4 (32 - 12 B^2 r^2 + 3 B^4 r^4) - M^2 (384 - 672 B^2 r^2 + 200 B^4 r^4 - 30 B^6 r^6) \nonumber \\{} & {} +\, M r (64 - 624 B^2 r^2 + 204 B^4 r^4 - 37 B^6 r^6).
    \label{eq:NEMS}
\end{align}

These are exact expressions wihout any approximation on $B$. All three fundamental frequencies remain nonzero, indicating stable oscillations in all directions. While $\Omega_\theta^{\text{EMS}}$ retains its Schwarzschild value due to dominant gravitational confinement in the polar direction, the azimuthal and radial frequencies are modified by magnetic curvature, altering the equatorial dynamics of the test particle.

Likewise, periastron and nodal precession frequencies are given by

\begin{align}
\Omega_{\text{per}}^{\text{EMS}} &= \frac{1}{\sqrt{4 - B^2 r^2}} \left[
\frac{(4 + B^2 r^2)^2}{16} \sqrt{ \frac{ M(4 - 3B^2 r^2) + 2B^2 r^3 }{ r^3 } } - \frac{1}{r^2 (4 + B^2 r^2)} \sqrt{N}
\right], \label{EMSper} \\
\Omega_{\text{nod}}^{\text{EMS}} &= \Omega_{\text{gL}} = \sqrt{\frac{M}{r^3}} \left[ \frac{(4+B^2 r^2)^2}{16} \sqrt{\frac{4-3B^2r^2+2B^2r^3}{4-B^2r^2}}  - 1 \right].
\end{align}

Remarkably, in the EMS spacetime, $\Omega_\phi^{\text{EMS}} \neq \Omega_\theta^{\text{EMS}}$, implying that the nodal precession frequency $\Omega_{\text{nod}}^{\text{EMS}}$ does not vanish even though the spacetime is nonrotating ($a \rightarrow 0$). This nonzero precession arises purely due to the curvature induced by the magnetic field. This magnetically induced precession — in the absence of spin — is known as the \textit{gravitational Larmor precession} (GLP: $\Omega_{\text{gL}}$) \cite{CCGLP}, a general relativistic analogue of classical Larmor precession that arises solely due to the presence of an external magnetic field.

\subsubsection{\textbf{Melvin's magnetic Universe ($a \rightarrow 0,M \rightarrow 0, B\neq0$)}}

In the Melvin magnetic universe $(a \rightarrow 0,M \rightarrow 0)$ \cite{melvinverse}, the exact fundamental precession frequencies are given by

\begin{align}
    \Omega_{\phi}^{\text{M}} &=  \frac{B\left(4+B^2 r^2\right)^2}{8 \sqrt{8-2 B^2 r^2}}, \label{eq:omegaphiMelvin} \\
    \Omega_{r}^{\text{M}} &= \frac{2 B}{4+B^2 r^2} \sqrt{\frac{32}{4-B^2 r^2}-3 B^2 r^2}, \label{eq:omegarMelvin} \\
    \Omega_{\theta}^{\text{M}} &= 0,
\end{align}

We observe a striking behavior in this purely magnetic spacetime: while the Keplerian frequency $\Omega_\phi^{\text{M}}$ and the radial epicyclic frequency $\Omega_r^{\text{M}}$ remain nonzero, the vertical epicyclic frequency $\Omega_\theta^{\text{M}}$ vanishes identically. This implies that although the geometry supports circular motion and radial oscillations, it offers no restoring force in the polar direction. A particle displaced from the equatorial plane does not oscillate back, indicating a complete breakdown of vertical stability.

Consequently, the nodal precession frequency formally becomes $\Omega_{\text{nod}}^{\text{M}} = \Omega_\phi^{\text{M}}$, but this does not correspond to a true precession of the orbital plane. There is no vertical deviation to precess around — a test particle cannot move off the equatorial plane and return. Thus, the term “nodal plane precession” becomes a misnomer: it arises purely from the definition, while physically it just reflects azimuthal motion in the absence of vertical dynamics. In contrast, the periastron precession frequency $\Omega_{\text{per}}^{\text{M}} = \Omega_\phi^{\text{M}} - \Omega_r^{\text{M}}$ remains physically meaningful, capturing genuine precession of the orbit’s periastron. The non-vanishing orbital precession in this spacetime arises solely from the magnetic energy density ($\sim B^2$), which contributes to spacetime curvature even in the absence of mass or spin \cite{Bonnor1960, CCGP}.

\subsubsection{\textbf{Slowly-rotating ($a_* \ll 1$) MKBH spacetime}}

Another important limiting regime of astrophysical relevance arises when a slowly rotating BH is immersed in a weak magnetic field. Specifically, we consider the slow-rotation approximation ($a_* \ll 1$) and assume the magnetic field strength is weak ($B \ll M^{-1}$). In this limit, the fundamental frequencies reduces to

\begin{align}
   \Omega_{\phi} \big|_{a\ll M} &\approx \sqrt{\frac{M}{r^3}}-\frac{a M}{r^3} + B^2 \left(\frac{r (M+r)}{4 \sqrt{M r}}-\frac{3 a M}{2 r}\right) ,
   \\
   \Omega_{r} \big|_{a\ll M} &\approx \frac{\sqrt{M (r - 6M)}}{r^2} +  \frac{3 a M (2 M + r)}{r^{7/2} \sqrt{r - 6M}} \nonumber
   \\
   & \quad + B^2 \left(\frac{ \left(6M^2 - 5Mr + r^2\right) }{ \sqrt{M}\sqrt{r - 6M}  } + \frac{3a (84 M^3-48 M^2 r+13 M r^2-3 r^3)}{4(r^2-6Mr)^{3/2} }  \right) ,
   \\
   \Omega_{\theta} \big|_{a\ll M} &\approx \sqrt{\frac{M}{r^3}} -\frac{3 a M}{r^3} - \frac{3}{4} a B^2 \left(1-\frac{3 M}{r}\right) ,
   \\
   \Omega_{\text{per}} \big|_{a\ll M} &\approx \frac{\sqrt{M r}-\sqrt{M (r-6 M)}}{r^2} -\frac{a M}{r^3}\left(1+\frac{3 (2 M+r)}{\sqrt{r (r-6 M)}}\right)+ B^2 N_3 ,
   \\
   \Omega_{\text{nod}} \big|_{a\ll M} &\approx \frac{2 a M}{r^3} + B^2 \left(\frac{3 a (r-5 M)}{4 r}+\frac{r (M+r)}{4 \sqrt{M r}}\right) ,
\end{align}

where

\begin{align}
    N_3 = \frac{\sqrt{M r}}{4} + \frac{r^2}{4 \sqrt{M r}} -\frac{\left(6 M^2-5 M r+r^2\right)}{\sqrt{M (r-6 M)}} 
    -\frac{3 a M}{2 r} +  \frac{3 a \left(-84 M^3+48 M^2 r-13 M r^2+3 r^3\right)}{4 (r (r-6 M))^{3/2}}.
\end{align}

These expressions reveal how even modest BH spin and magnetic field strength can introduce significant corrections to the orbital and epicyclic frequencies around a slowly-rotating weakly-magnetized Kerr BH. Importantly, no weak-gravity approximation was assumed. This makes the slow-rotation limit particularly relevant for modeling astrophysical systems such as QPOs, where BHs may rotate slowly but still exhibit strong-field behavior. The results thus serve as a useful framework for connecting theoretical predictions with astrophysical phenomena such as QPOs, where deviations from pure Kerr dynamics may become observationally significant.

\section{\label{sec:MagQPO}Effect of Magnetic Fields on Orbital Frequencies}

We now examine the effect of the external magnetic field on the orbital precession frequencies in MKBH spacetime. The analysis is restricted to circular geodesics lying within the domain bounded by the ISCO and OSCO, ensuring stability and astrophysical relevance. Of particular interest are the Keplerian frequency $\Omega_\phi$, the periastron precession frequency $\Omega_{\text{per}}$, and the nodal plane precession frequency $\Omega_{\text{nod}}$, which play a central role in the modeling of QPOs \cite{Stella1999, StellaLTQPO, StellaBHXB}. 

For our analysis, we consider magnetic field strengths of $B = 0.001M^{-1}$, $0.01M^{-1}$, and $0.1M^{-1}$, spanning a range relevant for compact object environments—from weakly magnetized systems to highly magnetized regimes. To frame these values in a physical context, consider a  SMBH of mass $M \sim 10^9 M_\odot$: a magnetic field strength of $B = 0.01 M^{-1}$ corresponds to a physical magnetic field of approximately $10^8$ G, which is considered strong by astrophysical standards. This validates our choice of parameter range for probing magnetic field effects.

\subsection{\label{subsec:OmegaPhi}Effect of Magnetic Field on $\Omega_\phi$}

We examine $\Omega_{\phi}$ as a function of $r$ for different spin parameters $a_*$ and magnetic field strengths $B = 0.001$, $0.01$, and $0.1 \, M^{-1}$. This setup enables a systematic analysis of how varying magnetic field strengths—from weak to strong—affect the Keplerian orbital frequency.

\begin{figure}[H]
\begin{center}
\subfigure[$B=0.001 \ M^{-1}$]{
\includegraphics[width=2.7in,angle=0]{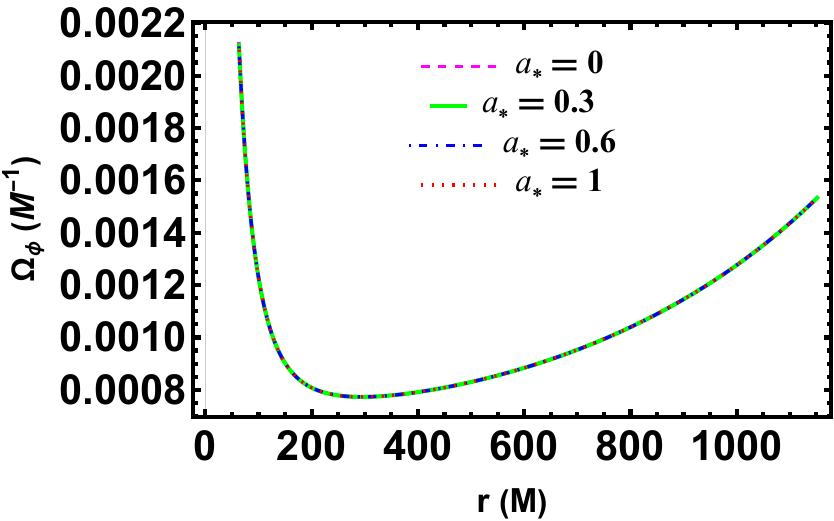}}
\subfigure[$B=0.01 \ M^{-1}$]{
\includegraphics[width=2.7in,angle=0]{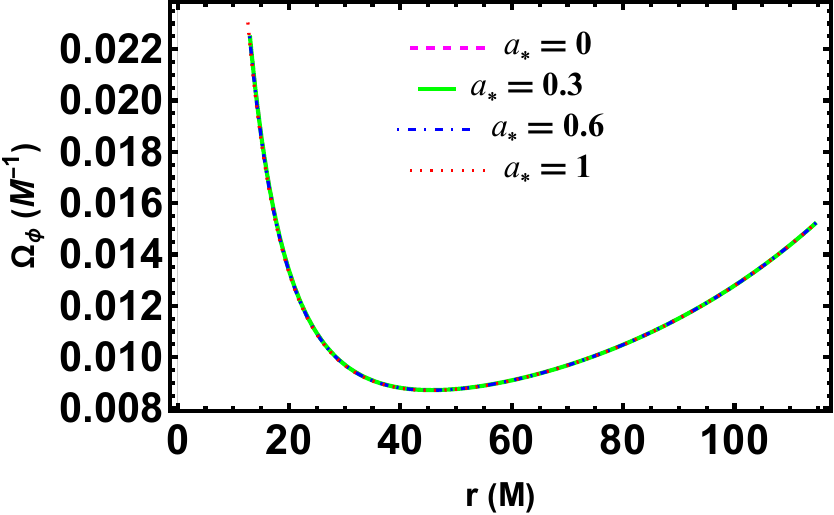}}
\subfigure[$B=0.1 \ M^{-1}$]{
\includegraphics[width=2.7in,angle=0]{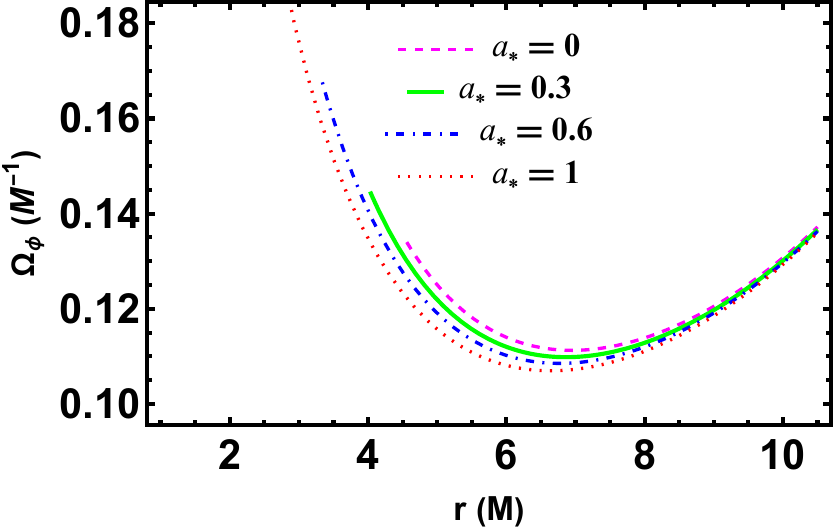}}
\caption{\label{Fig:Omegaphi} Radial profile of $\Omega_\phi$ (in $M^{-1}$) for a MKBH, showing non-monotonic behavior due to the competing effects of spacetime curvature and magnetic field. Results are plotted from $r_{\text{ISCO}}$ to $r_{\text{OSCO}}$ for different values of $a_*$. See Sec.\ref{subsec:OmegaPhi} for details.}
\end{center}
\end{figure}

The Keplerian frequency $\Omega_\phi$ initially increases near the BH, then decreases at intermediate radii, and rises again at larger distances. Close to the BH, strong spacetime curvature dominates and enhances $\Omega_\phi$. At intermediate radii, the competing influences of gravity and magnetic field produce a local minimum in the frequency profile. Beyond this region, the magnetic field becomes dominant, leading to an increase in $\Omega_\phi$. This nontrivial behavior—particularly the upturn at larger radii—demonstrates that the magnetic field supports orbital motion and contributes positively to the Kepler frequency. Physically, the magnetic field acts as a confining force at large radii, effectively supporting orbital motion even far from the BH. This magnetically induced upturn is a feature absent in asymptotically flat spacetimes, making it a distinct hallmark of the MKBH geometry. 
The overall trend is clearly illustrated in Fig.~\ref{Fig:Omegaphi}.

\subsection{\label{subsec:OmegaNod}Effect of Magnetic Field on $\Omega_\text{nod}$}

\begin{figure}[h!]
\begin{center}
\subfigure[$B=0.001 \ M^{-1}$]{
\includegraphics[width=2.7in,angle=0]{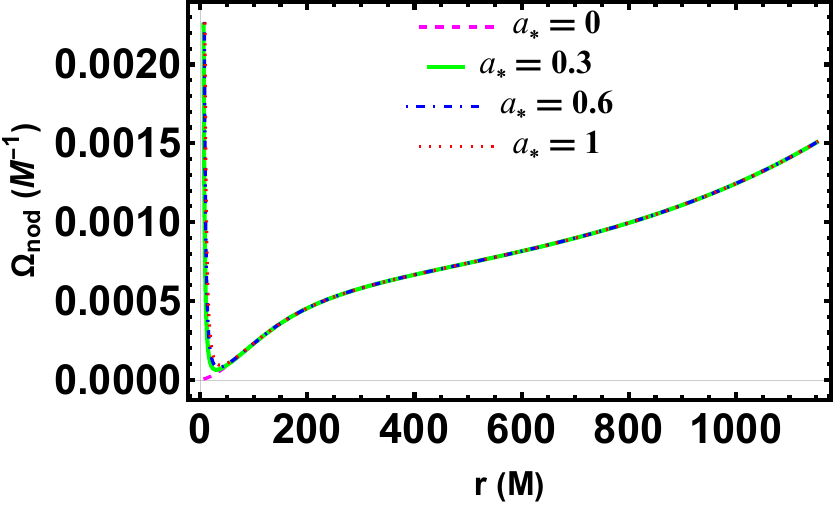}}
\subfigure[$B=0.01 \ M^{-1}$]{
\includegraphics[width=2.7in,angle=0]{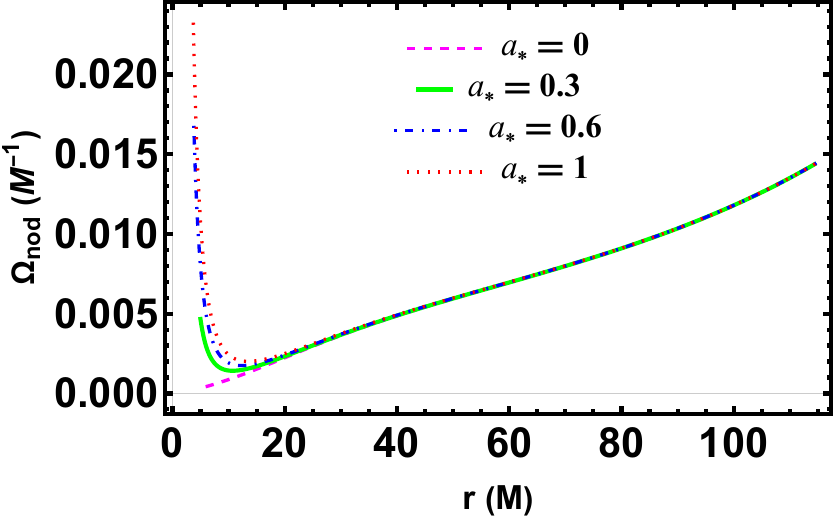}}
\subfigure[$B=0.1 \ M^{-1}$]{
\includegraphics[width=2.7in,angle=0]{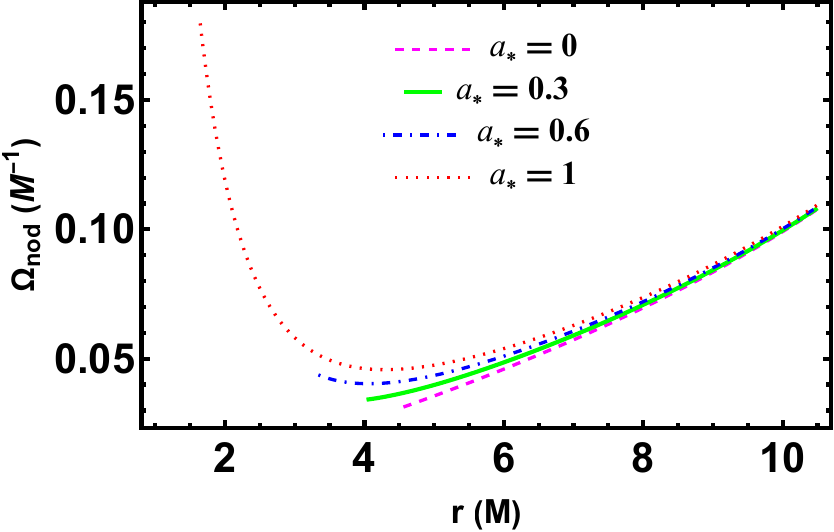}}
\caption{\label{Fig:Omeganod} Radial profile of $\Omega_{\text{nod}}$ (in $M^{-1}$) for a MKBH, showing non-monotonic behavior due to competing frame-dragging and magnetic effects. Results are plotted from $r_{\text{ISCO}}$ to $r_{\text{OSCO}}$ for different values of $a_*$ and $B$. See Sec.\ref{subsec:OmegaNod} for details.}
\end{center}
\end{figure}

The nodal precession frequency $\Omega_\text{nod}$ also exhibits a characteristic non-monotonic dependence on radial distance $r$. Near the BH, $\Omega_\text{nod}$ increases due to strong frame-dragging (Lense-Thirring) effects. At intermediate radii, a competition between frame-dragging and the magnetic field leads to a local minimum in the profile. At larger radii, the magnetic energy becomes dominant, causing $\Omega_\text{nod}$ to rise again—an imprint of magnetically dominated precession. Here too, the magnetic field acts as a long-range confining agent, sustaining precession frequencies that would otherwise decay in an asymptotically flat Kerr spacetime. This behavior is clearly visible in Fig.~\ref{Fig:Omeganod}.

In the non-rotating case ($a_* = 0$), LT precession vanishes, and $\Omega_\text{nod}$ reflects purely magnetic effects, giving rise to GLP without the divergence near the BH caused by frame-dragging. This aligns with the findings of Ref.\cite{CCGLP}. As the magnetic field strength increases (e.g., $B = 0.1 M^{-1}$), magnetic energy dominates even at small radii, overwhelming gravitomagnetic contributions — with extremal BHs constituting the sole exception. Here, the gravitational Meissner effect expels poloidal magnetic energy from the near-BH region, allowing frame-dragging to maintain dominance even in ultra-strong magnetic environments (see Panel c of Fig.~\ref{Fig:Omeganod}).

\subsection{\label{subsec:OmegaPer}Effect of Magnetic Field on $\Omega_\text{per}$}

\begin{figure}[h!]
\begin{center}
\subfigure[$B=0.001 \ M^{-1}$]{
\includegraphics[width=2.7in,angle=0]{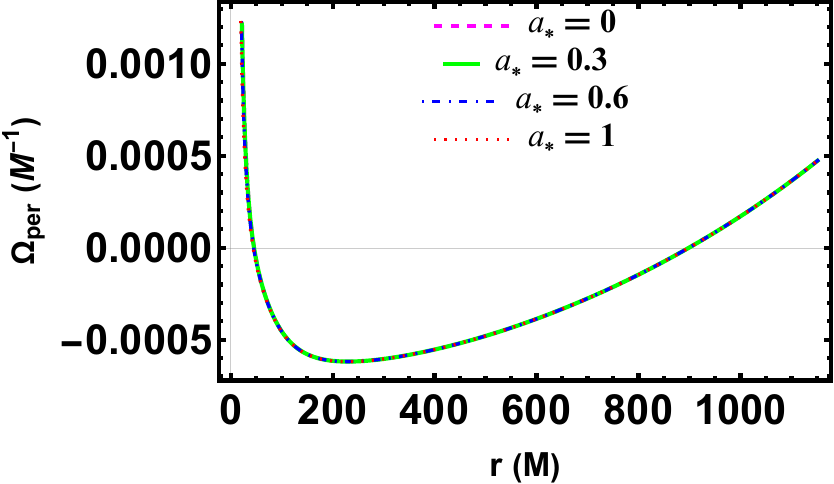}}
\subfigure[$B=0.01 \ M^{-1}$]{
\includegraphics[width=2.7in,angle=0]{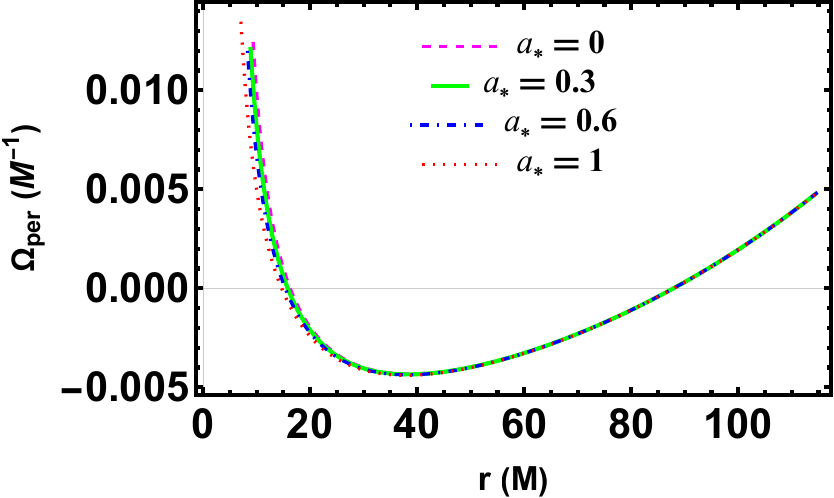}}
\subfigure[$B=0.1 \ M^{-1}$]{
\includegraphics[width=2.7in,angle=0]{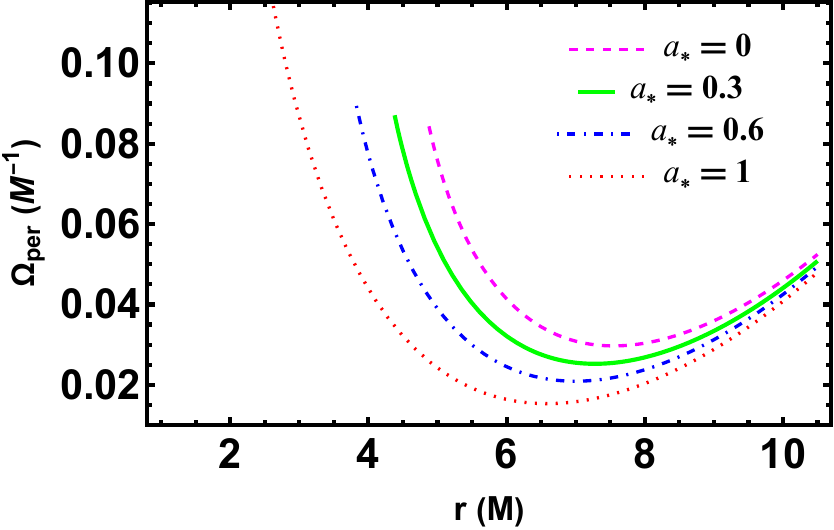}}
\caption{\label{Fig:Omegaper} Radial profiles of $\Omega_{\text{per}}$ (in $M^{-1}$) for a MKBH, plotted from $r_{\text{ISCO}}$ to $r_{\text{OSCO}}$. For $B \leq B_\text{neg}$ [panels (a), (b)], a retrograde precession window appears where $\Omega_{\text{per}}<0$. For $B > B_\text{neg}$ [panel (c)], the retrograde branch vanishes and $\Omega_{\text{per}}$ remains strictly positive throughout. See Sec.\ref{subsec:OmegaPer} for details.}
\end{center}
\end{figure}

The periastron precession frequency, $\Omega_\text{per}$, exhibits distinct radial dependence in the presence of an external magnetic field. Close to the BH, strong curvature effects dominate, and $\Omega_\text{per}$ is large and positive. At larger radii, the $B^2$–induced curvature contribution becomes increasingly important, subtly modifying the precession dynamics across the stability domain. This interplay is evident in Fig.~\ref{Fig:Omegaper}. For sufficiently weak magnetic fields ($0 < B < 0.05 M^{-1}$), $\Omega_\text{per}$ turns negative within a finite radial band [panels (a) and (b)], signalling a retrograde precession of the orbit. Intriguingly, this retrograde window disappears and the precession remains strictly positive [panel (c)] for comparatively stronger magnetic fields. In the realistic astrophysical scenarios, the value of $B$ is supposed to be much less than $B < 0.05 M^{-1}$. As evident from panels (a) and (b) of Fig.~\ref{Fig:Omegaper}, the negative precession window hardly depend on the value of spin parameter $a_*$. This is expected, as the effect of $a_*$ decreases significantly far from the event horizon, unlike $B$. Thus, let us analyze the negative precession in the light of non-rotating EMS BH, which is the simplest case as well.

To understand the negative $\Omega_{\rm per}$, let us first determine the negative precession window, i.e., the radii of the orbits where $\Omega_{\rm per}$ vanishes. Considering the non–rotating EMS BH, we expand the analytical expression for $\Omega_\text{per}^{\rm EMS}$ (Eq. \ref{EMSper}) at large radii ($M/r \ll 1$). The leading term and $B^2$ pieces balance at a radius

\begin{align}
    r_{n1} \sim \sqrt{\frac{2 M}{B}}.
\end{align}

The outer crossover radius $r_{n2}$ is determined by the condition $\Omega_{\text{per}} = 0$, which marks the transition between prograde and retrograde precession. This crossover occurs at large radii, $r \sim 1/B$, where the magnetic curvature dominates over the gravitational contribution. In this regime, the spacetime is well approximated by the Melvin background. Accordingly, the crossover condition can be obtained analytically by setting $\Omega_{\text{per}}^{\text{M}} = \Omega_\phi^{\text{M}} - \Omega_r^{\text{M}} = 0$ (see Eqs.~\ref{eq:omegaphiMelvin} and \ref{eq:omegarMelvin} and the discussion below). Solving this equation for $r \equiv r_{n2}$ yields
\begin{align}
    r_{n2} \approx \frac{0.9}{B}.
\end{align}
This scaling follows from the Melvin approximation and provides a reliable analytic estimate for the outer crossover radius in the magnetically dominated regime. Consequently, the negative–precession interval is bounded from below by $r_{n1} \propto 1/\sqrt{B}$ and from above by $r_{n2} \propto 1/B$. Moreover, this inner crossover lies close to the ISCO and is only physically meaningful if it occurs outside the innermost stable orbit. In the Schwarzschild limit ($B=0$) the ISCO is at $6M$, but in the magnetized case the ISCO radius decreases from $6M$ toward $2M$ as $B$ grows, before eventually disappearing inside the horizon. For weak magnetic fields ($B \ll M^{-1}$), the ISCO remains close to $6M$, so requiring $r_{n1} > r_\text{ISCO}(B)$ implies a critical bound
\begin{align}
    B < \frac{1}{18M} \approx 0.055 \ M^{-1}=B_\text{neg},
\end{align}
below which a negative precession branch exists. Above this threshold the crossover drops beneath the ISCO, and the negative branch is no longer physically accessible. A similar conclusion holds for the MKBH case, confirming that the disappearance of retrograde precession at higher magnetic field strengths is a robust feature of the geometry.

\begin{table}[h!]
\centering
\setlength{\tabcolsep}{8pt} 
\renewcommand{\arraystretch}{1.3}
\begin{tabular}{|c|c|c|c|c|c|c|c|c|c|c|c|}
\hline
$a_*$ & 0.0 & 0.1 & 0.2 & 0.3 & 0.4 & 0.5 & 0.6 & 0.7 & 0.8 & 0.9 & 1.0 \\
\hline
$B_\text{neg} \ (M^{-1})$ &
0.055 & 0.056 & 0.057 & 0.058 & 0.059 & 0.060 & 0.061 & 0.062 & 0.063 & 0.065 & 0.067 \\
\hline
\end{tabular}
\caption{Threshold magnetic field values $B_\text{neg}$ for different spin parameters $a_*$, below which a negative-frequency window in the range $r_{n1} < r < r_{n2}$ appears.}
\label{tab:Bneg}
\end{table} 

Table~\ref{tab:Bneg} summarizes the values of $B_\text{neg}$ across the full spin ($a_*$) range, which indicates that the change in the value of $B_\text{neg}$ increases maximum $\sim 20\%$ for the extremal MKBH. For an EMS BH ($a_*=0$), retrograde precession occurs for $B \leq 0.055  M^{-1}$, while for an extremal MKBH ($a_*=1$), the threshold increases slightly to $B \leq 0.067 M^{-1}$. Increasing spin therefore allows the retrograde window to persist at moderately stronger magnetic fields. Furthermore, for very weak fields ($B \ll B_\text{neg}$), the $r_{\text{OSCO}}$ shifts outward, and the retrograde branch appears at progressively larger orbital radii. 

The emergence and disappearance of retrograde precession highlight the nontrivial interplay between gravitational and magnetic curvature. Although the test particle is electrically neutral, spacetime curvature sourced by the magnetic energy density ($\sim B^2$) modifies geodesic structure in such a way that periastron dynamics are fundamentally altered. This effect provides a striking example of how magnetic fields can leave observable imprints on orbital motion, even without direct electromagnetic coupling. However, for $B \rightarrow 0$, both $r_{n1}$ and $r_{n2}$ tend to infinity. Thus, the negative periastron precession could not be realized in case of the non-magnetized Schwarzschild and Kerr BH spacetime.

Realistic astrophysical BHs are expected to be immersed in a comparatively weak ($B \ll B_{\rm max}$) but non-zero magnetic fields, as discussed in Sec.~\ref{sec:Intro}. Thus, in principle, the negative periastron precession could be realized at some finite distance from such a BH. From an astrophysical perspective, the potential for precessional reversal is intriguing. The influence of an ambient, large-scale magnetic field on orbital precession can extend far beyond the sphere of gravitational dominance of a central BH. To contextualize this within an astrophysical setting, let us consider the active binary system Cygnus X-1, which hosts a stellar-mass BH of mass $M \approx 14\,M_\odot$~\cite{Orosz2011}. As a crude estimation, we assume that the BH is immersed in a uniform magnetic field of strength $B \sim 10^3\,\text{G}$~\cite{Zanin2016}, corresponding to $B \sim  10^{-16}\,M^{-1}$, extending up to a distance of $\sim 10^{8}\,\text{km}$ \cite{Zanin2016} from the BH.
This relatively weak field nonetheless generates a finite domain for retrograde periastron precession. The inner boundary of the negative precession window is given by $r_{n1} \sim 10^{8} M$ in this case, which corresponds to a physical distance of $\sim 6$ AU. In contrast, the outer boundary, scaling as $r_{n2} \sim 1/B$, is of order $10^{15} M$ ($\sim 1$ kpc), rendering it astronomically vast and observationally irrelevant, since the kG order magnetic fields do not extend to such distances. The key result is that $r_{n1}$ lies within a physically meaningful region. This implies that a test particle on a bound orbit around Cygnus X-1 could in principle experience a crossover into a regime of magnetically-induced negative periastron precession as it moves inward past this radial threshold, although the specific frequency and amplitude of this precession would require further analysis. Additional research in this area is much needed to clarify the implications in more detail.

\section{\label{sec:QPOImplications}Astrophysical Implications and QPO Modeling in Magnetized Kerr Spacetime}

QPOs are among the most informative timing features observed in black hole X-ray binaries (BHXBs), appearing as narrow peaks in the X-ray power spectrum \cite{belloni2014fast}. They are typically grouped into two classes: high-frequency QPOs (HFQPOs), in the range of tens to a few hundred Hz, and low-frequency QPOs (LFQPOs), spanning $0.01-30$ Hz. Certain BHXBs, such as XTE J1650–500 and 4U 1630–47, have shown HFQPOs in the ranges $50-270$ Hz and $150-450$ Hz, respectively \cite{Belloni2012}, sometimes with multiple peaks in harmonic ratios—strongly suggesting an origin in coherent orbital dynamics close to the BH.

LFQPOs are further divided into three types (A, B, and C) based on their spectral–timing properties, with typical centroid frequencies around $6.5-8$ Hz (type A), $0.8-6.4$ Hz (type B), and $0.01-30$ Hz (type C) \cite{CCDKNS}. Among the various theoretical models proposed, the relativistic precession (RP) model \cite{StellaLTQPO} offers a particularly direct physical interpretation: it connects QPOs to the fundamental frequencies of geodesic motion in strong gravity. In this framework, the orbital frequency $\Omega_\phi$, the radial epicyclic frequency $\Omega_r$, and the vertical epicyclic frequency $\Omega_\theta$ generate observable combinations such as the periastron precession frequency $\Omega_{\text{per}} = \Omega_\phi - \Omega_r$ and the nodal plane precession frequency $\Omega_{\text{nod}} = \Omega_\phi - \Omega_\theta$ \cite{StellaLTQPO,StellaBHXB,IngramRP}. Applied to BHXBs, the RPM associates the upper HFQPO with $\Omega_\phi$, the lower HFQPO with $\Omega_{\text{per}}$, and the C-type LFQPO with $\Omega_{\text{nod}}$ \cite{Motta2014a,Motta2014b}, thereby enabling mass and spin constraints from timing data alone.

Most BHXBs are transient systems, where accretion disks form only during outbursts. Even in persistent BHXBs, the accretion flow undergoes state transitions, implying variations in disk structure and truncation radius. Consequently, the inner edge of the disk may sometimes advance toward the BH and sometimes recede, depending on accretion rate and spectral state. If QPOs are indeed connected to the natural frequencies described above, then such disk dynamics should be reflected in their evolution, since the frequencies vary sensitively with radial distance. Observationally, systematic drifts in QPO frequencies across state transitions are indeed reported \cite{Motta2014a,Motta2014b}.

In the MKBH scenario, the presence of the magnetic field modifies the radial profiles of $\Omega_{\phi,r,\theta}$ and $\Omega_{\text{nod}}$, and can even induce retrograde precession in $\Omega_{\text{per}}$ under suitable conditions. Such departures from Kerr dynamics could, in principle, be uncovered through careful tracking of QPO frequency evolution as the disk moves inward or outward.

Complementary probes such as relativistic Fe K$\alpha$ line profiles and X-ray polarization, both sensitive to spacetime geometry and precession dynamics, can provide independent tests of the MKBH framework \cite{CCDKNS, CCGMmonople}. Multi-messenger timing–spectral strategies of this kind may help disentangle magnetic corrections from spin degeneracies and turn the present theoretical results into observationally falsifiable predictions.
\\

The magnetized Kerr geometry is not asymptotically flat, since the external magnetic field extends to infinity. At first sight, this raises the concern of whether orbital frequencies derived in such a background can be meaningfully compared with QPO observations, which are usually interpreted in asymptotically flat settings. However, several points justify this approach. First, the dynamics relevant to QPO generation arise in the inner accretion flow, typically between the ISCO and a few gravitational radii from the BH. In this near-horizon region, the geometry is well-approximated by Kerr with $\mathcal{O}(B^2)$ corrections due to the magnetic field \cite{Taylor2025}, while the Melvin-type asymptotics appear only at very large radii $r \sim 1/B$ \cite{Taylor2025}, far outside the QPO region. Second, in magnetized spacetimes the domain of stable circular motion is naturally bounded between two radii: the ISCO at the inner edge and the OSCO at the outer edge. This confinement, which arises directly from the magnetic modifications to the geometry, ensures that all physically relevant orbital frequencies are defined within a finite and astrophysically meaningful interval [$r_{\text{ISCO}}, r_{\text{OSCO}}$], without invoking the pathological region at infinity. Moreover, as emphasized in \cite{Taylor2025}, the Ernst–Wild geometry provides the only exact analytic solution that fully incorporates the backreaction of the magnetic field on the black hole spacetime, thereby offering a consistent framework beyond test-field models like Wald’s solution\cite{WaldBHmag}. In addition, the study of perturbations in magnetized BH backgrounds reveals qualitatively new dynamical effects—such as mode confinement and superradiant instabilities—that rely critically on the full magnetized asymptotics and cannot be captured by asymptotically flat approximations \cite{Brito2014}. These phenomena set intrinsic limits on BH spin evolution and demonstrate that the global structure of the spacetime, even if irrelevant for local disk dynamics, plays a key role in establishing the physical consistency of the model. Finally, in realistic systems the magnetic fields are much weaker than the characteristic scale $B_{\rm max} \sim 1/M$, so that the OSCO lies at radii far beyond those probed by the inner accretion disk. In this regime, the non-flat asymptotics are irrelevant for QPO modeling, and the MKBH spacetime provides a consistent and physically motivated framework for connecting fundamental frequencies to observations.

\section{\label{sec:dis} Summary and Discussion}

In this work, we have carried out a detailed analysis of the equatorial geodesic structure and the orbital frequency spectrum of a Kerr black hole immersed in an external magnetic field. Our analysis highlights the role of magnetic curvature in shaping both null and timelike geodesics, thereby offering new insights into strong-field dynamics in non-asymptotically flat spacetimes. The main results of this study can be summarized as follows:

\begin{enumerate}
    \item Starting from the basic formulation, we have derived the potential function governing the equatorial timelike geodesics in the full magnetized Kerr spacetime. The magnetic field introduces a substantial modification, manifesting as a potential barrier that grows with $B$ similar to \cite{BHshield}. However, the effect of $a_*$  is comparatively small on the potential barrier.

    \item By determining the radius of CPO, we have identified a critical magnetic field strength $B_{\text{cr}}(a_*)$, beyond which no circular geodesics—timelike or null—can exist.

    \item For sub-critical fields ($B < B_{\text{cr}}$), the CPO equation yields two positive roots. We propose a novel identification of the outer root as the \textit{outermost stable circular orbit}, providing a physically motivated definition of the OSCO based on the null geodesic structure. 
    
    \item We have obtained exact analytical expressions for the fundamental orbital frequencies--Keplerian, radial, and vertical epicyclic--in the MKBH spacetime. For conciseness, we have presented here their series expansions up to $\mathcal{O}(B^2)$ term, which are directly relevant for the astrophysical observations of (non-)rotating BHs immersed in magnetic fields.  In the slow-rotation and weak-magnetic field regime ($a/M \ll 1$, $B \ll M^{-1}$), our expressions yield tractable forms, demonstrating that even modest spin and magnetic fields produce appreciable corrections in the strong-field region.
    These expressions clearly demonstrate how magnetic curvature alters the relativistic precession characteristics of particle orbits.

    \item We have derived the exact expressions for the ISCO radius and OSCO radius, which are applicable to any arbitrary values of $a_*$ and $B$. We have shown that the both radii decrease monotonically with increasing $B$ for the whole range of $a_*$: $0 \leq a_* < 1$. In the extremal BH case ($a_* \rightarrow 1$), the ISCO radius remains fixed at the horizon ($r/M \rightarrow 1$)  for any value of $B$ due to the gravitational Meissner effect, unlike OSCO. Interestingly, Fig. \ref{Fig:CPO} shows that ISCO, OSCO and CPO all three curves coincide at a particular point $P$ which gives rise to $B_{\rm cr}$.

    \item The Keplerian frequency $\Omega_\phi$ and the nodal plane precession frequency $\Omega_\text{nod}$ exhibit a similar non-monotonic trend. Close to the BH, frame dragging effect due to $a_*$ dominates, whereas at larger radii magnetic curvature takes over similar to \cite{CCGLP}. In the non-rotating case ($a_*=0$), nodal plane precession is entirely magnetic in origin, consistent with gravitational Larmor precession \cite{CCGLP}.

    \item We have reported a reversal to retrograde periastron precession ($\Omega_\text{per} < 0$), a phenomenon absent in the standard Kerr and Schwarzschild spacetime. For the static case, we have identified the condition for this reversal, which defines a critical magnetic field strength $B_\text{neg}$ above which the precession is strictly prograde. This behavior generalizes to rotating MKBH, with the critical value $B_\text{neg}$ increasing modestly with  $a_*$. A crude estimation reveals that such a negative precession could be realizable in case of a realistic astrophysical BH immersed even in a weak magnetic field.

\end{enumerate}

The theoretical results outlined above have direct implications for the astrophysical modeling of BHs immersed in magnetic environments. Although magnetic field strengths near BHs remain highly uncertain, estimates span a wide range: values as high as $10^7$ G have been reported in the corona of Cygnus X-1 \cite{santomnrs}, while significantly weaker fields—$30$–$100$ G for Sgr A* \cite{eatough2013strong} and $1$–$30$ G for M87* \cite{EHT7,EHT8}—are inferred from radio and polarization measurements. Ultra-strong fields comparable to Eq.~\eqref{bmax} have not been observed and may be suppressed by magnetic shielding effect \cite{BHshield}, though extreme cases remain plausible. For instance, magnetar companions (e.g. SGR J1745–29 near Sgr A* with $B \sim 1.6 \times 10^{14}$ G \cite{kennea2013swift,mcgill}) or primordial relic fields up to $10^{20}$ G \cite{GRASSO1,GRASSO2} could immerse primordial BHs in intense backgrounds. Additionally, our formulation could also be applicable to BHs formed through the merger of a BH with one or more magnetized neutron stars, magnetars \cite{lyu83, east, BHshield}, or magnetized stellar progenitors. In such cases, the magnetic field may remain anchored to the newly formed BH rather than sliding-off, as shown in \cite{lyu, lyu83} (though see \cite{brp}).
In such regimes, geodesic motion may be strongly restricted, potentially suppressing accretion and stellar orbits \cite{BHshield}, rendering BHs elusive to conventional probes. Their presence could still be inferred through lensing of background sources, while small oscillations around circular orbits remain informative: epicyclic modes may persist even below the shielding threshold and generate detectable QPOs.

Against this backdrop, a key result of our analysis is the magnetically induced reversal of the periastron precession frequency $\Omega_{\text{per}}$. Such behavior may leave distinctive signatures: HFQPO phenomenology, commonly interpreted via the RPM \cite{StellaLTQPO,Stella1999} where $\Omega_{\text{per}}$ is associated with lower HFQPOs in GRO J1655–40 and XTE J1550–564 \cite{Remillard2006}, could be modified once magnetic fields are included. Our results thus open a new avenue for interpreting QPOs in magnetized flows, complementing earlier magnetic corrections in literature \cite{CCPenrose,Karthik1,HuoKerrMelvin}. Moreover, retrograde precession itself may be observable: stellar orbital precession measured by GRAVITY \cite{GRAVITY2020} and SINFONI \cite{Eisenhauer2005} already constrains spacetime geometry, and upcoming high-precision facilities such as BHEX \cite{BHEX} and ngEHT \cite{ngEHT} could directly probe these effects near BHs.

The presence of an outermost stable circular orbit introduces a qualitatively new restriction on bound motion around MKBH. If, for instance, a neutron star orbits a BH at a radius larger than $r_{\text{OSCO}}$, the orbit cannot remain stable under perturbations. In such a case, even small deviations would drive the neutron star away from a circular trajectory, leading either to its ejection from the system or to a rapid inspiral toward smaller radii within the stable band $[r_{\text{ISCO}}, r_{\text{OSCO}}]$. This has direct astrophysical implications: binary configurations involving compact companions are viable only within this finite radial window, while accretion disks fed by such companions would naturally truncate at $r_{\text{OSCO}}$. Thus, the OSCO acts as a dynamical outer boundary, shaping both the long-term survival of binaries and the radial extent of matter distribution around magnetized BHs. However, it is important to note that the OSCO occurs at radii $r \sim \mathcal{O}(1/B)$, where the spacetime enters the magnetically 
dominated (Melvin-like) regime. In this region, the assumption of a 
uniform external magnetic field becomes increasingly idealized and may 
not accurately represent realistic astrophysical environments. 
Accordingly, while the OSCO defines a well-posed outer stability boundary 
within the present framework, its direct identification with physical 
truncation radii (e.g., disk edges or binary stability limits) should be 
treated with caution. Rather, it is more appropriate to interpret the 
OSCO as a feature of the underlying idealized geometry, whose qualitative 
implications may inform, but not directly determine, astrophysical systems.

Our results may inform extensions of the Bardeen–Petterson (BP) alignment framework \cite{BPOG}. The non-monotonic radial behavior of the nodal precession frequency $\Omega_{\text{nod}}$ suggests that the classical warp radius would acquire a magnetic-field dependence. In this view, warped disks would be confined to a finite radial domain $[r_{\text{ISCO}}, r_{\text{OSCO}}]$, with boundaries shaped jointly by $a_*$ and $B$. Such magnetic corrections could complement GRMHD studies of warped structures in magnetically arrested disks \cite{Liska2019,Teixeira2014,Tchekhovskoy2011}, could be relevant for X-ray binaries \cite{IngramRP,Banerjee2019, banerjee2019n}, endoparasitic BH inside a strongly-magnetized host star \cite{adarsha2, adarsha2025,  CCSubsolar, CBJDarkcore} and tidal disruption events \cite{Stone2019}. Interestingly, recent observational work has uncovered direct signatures of Lense–Thirring precession in highly magnetized environments \cite{Farah2025LT}. A superluminous supernova was reported to be powered by a magnetar whose tilted fallback disk undergoes LT precession, producing quasi-periodic modulations in the light curve—the first observational evidence of frame-dragging–induced precession in a strongly magnetized compact object. Our results, which demonstrate how external magnetic curvature reshapes $\Omega_{\text{nod}}$ in Kerr spacetimes, provide a theoretical complement by highlighting how magnetic fields can modulate precession frequencies in such cases.

The magnetized Kerr spacetime employed in our study is indeed an exact solution to the Einstein–Maxwell equations but is not asymptotically flat. This arises from the assumption of an asymptotically uniform magnetic field, a widely adopted simplification in the literature \cite{Karthik1}. Though idealized, such configurations are analytically tractable and have successfully modeled various astrophysical phenomena involving magnetized BHs (see \cite{Dadhich2018, CCFaraday, CCPenrose, Chatterjee2017, Magkerrmeissner} and references therein). Given the current lack of precise measurements of magnetic field geometries near compact objects \cite{punslyBHGH, Davidengine}, this assumption remains justified. While mathematically consistent, this setup does not result from realistic astrophysical collapse and leads to a spacetime that asymptotically approaches the Melvin universe, limiting the direct physical interpretability of all astrophysical observables. Nonetheless, such models are valuable for isolating the qualitative effects of external magnetic fields on neutral test particles in strong gravity regimes. Here, our aim is not to construct a fully realistic model but to uncover trends that may persist in more complete treatments. Future refinements could include spatially varying magnetic fields or self-consistent field sources, potentially through numerical relativity. Our exact formulation can be appreciated from a more mathematical perspective, highlighting certain intriguing features of the spacetime. While these characteristics are mathematically consistent within the Einstein-Maxwell framework, they restrict the direct applicability and physical interpretation of the results in realistic astrophysical scenarios.

Finally, we have shown that the exact magnetized Kerr spacetime gives rise to global modifications of equatorial geodesic dynamics that have no counterpart in the Kerr geometry. Although the radial coordinate serves only as a parameter within a chosen representation, the principal quantities examined here --- the orbital and epicyclic frequencies and the associated precession rates --- are physical, gauge-invariant observables directly connected to measurable timing phenomena such as QPOs \cite{StellaLTQPO,StellaBHXB,IngramRP}. In contrast to our earlier work \cite{Karthik1}, which dealt with spin-induced precession along nongeodesic trajectories \cite{StraumannGR, CCDKNS, Chatterjee2017}, the present study focuses on the global structure, stability boundaries, and dynamical properties of geodesic motion itself. These results therefore establish new geometric and dynamical features of particle motion in the Einstein--Maxwell magnetized Kerr spacetime with potential observational implications.
\\

{\bf Acknowledgements:}
The authors acknowledge the support of Manipal Academy of Higher Education. The authors also thank the referee for constructive comments that helped improve the clarity of this paper.

\appendix

\section{\label{app:Vefflimits}Limiting Cases of the Potential Function}

To validate the consistency of our expression for the potential function derived in Eq.~\eqref{eq:VeffMKBH}, we examine three physically relevant limiting cases: (i) the Kerr spacetime in the absence of a magnetic field, (ii) the magnetized Schwarzschild spacetime (i.e., $a \rightarrow 0$), and (iii) the Melvin universe limit.

\subsection{Kerr Limit ($B=0$)}

When the magnetic field is switched off, the MKBH spacetime reduces to the standard Kerr spacetime. In this limit, the magnetization functions reduce to unity, $\Lambda_e = \Lambda_0 = 1$, and $A_e = (r^2+a^2)^2 - a^2 \Delta$. Consequently, Eq.~\eqref{eq:VeffMKBH} simplify to the well-known $V_\mathrm{r} (r)$ for equatorial timelike geodesics in the Kerr geometry \cite{ferrari2020GR}:

\begin{align}
   V_r (r) \big|_{B \to 0} = \frac{1}{r^4} \left[ (r^2+a^2)^2 - a^2 \Delta \right] (E-V_+) (E-V_-) - \frac{\Delta}{r^2},
\end{align}
with the potential functions given by
\begin{align}
    V_\pm = \frac{-2aMr \pm r^2 \sqrt{\Delta}}{(r^2+a^2)^2 - a^2 \Delta} L,
\end{align}
which is in agreement with Ref.~\cite{ferrari2020GR}.

\subsection{Magnetized Schwarzschild Limit ($a = 0$)}

For $a \rightarrow 0$, the MKBH metric reduces to the magnetized Schwarzschild solution. In this case, $\varpi_e|_{a \rightarrow 0}=0$, $\Lambda_e|_{a \rightarrow 0} = 1 + \tfrac{1}{4} B^2 r^2 \equiv \Lambda_{\text{M}}$, and $\Delta|_{a \rightarrow 0} = r^2 - 2M r \equiv \Delta_{\text{Sch}}$. In the Schwarzschild limit we find

\begin{align}
    V_+ + V_- \propto a = 0, \quad V_+ V_- = - \frac{L^2 \Delta_{\text{Sch}} \Lambda_M^2}{r^4}
\end{align}

such that

\begin{align}
(E - V_+) (E - V_-) 
&= E^2 - \frac{L^2 \Delta_{\text{Sch}} \Lambda_M^2}{r^4}
\end{align}

Substituting into Eq.~\eqref{eq:VeffMKBH} yields

\begin{align}
    \Lambda_M^4 \dot{r}^2 = E^2 - V_\text{eff}
\end{align}

where we identify the effective potential as \cite{Galtsov1978,BHshield}

\begin{align}
    V_\text{eff} = \frac{\Delta_\text{Sch} \Lambda_M^2}{r^2} \left(1 + \frac{L^2 \Lambda_M^2}{r^2} \right)
\end{align}

which recovers the known result for magnetized Schwarzschild spacetime~\cite{Galtsov1978}.

\subsection{Melvin Universe Limit ($M \rightarrow 0$, $a \rightarrow 0$)}

If both the mass and spin vanish, the geometry reduces to the pure Melvin universe, which is static, axisymmetric, and sourced entirely by a magnetic field. Following analogous steps to the magnetized Schwarzschild case, in this limit the effective potential simplifies to \cite{YKLimMelvin}

\begin{align}
V_{\mathrm{eff}} = \frac{\Lambda_{\text{M}}^4 L^2}{r^2} + \Lambda_{\text{M}}^2.
\end{align}

\section{\label{app:general} Fundamental Precession Frequencies in General Stationary and Axisymmetric Spacetimes}

Here, we summarize the derivation of orbital and epicyclic frequencies for a general stationary and axisymmetric spacetime \cite{Ryan1995}. We begin with the standard normalization condition for the four-velocity $u^\mu = dx^\mu/d\tau$ of a test particle:

\begin{align}
    -1 = g_{tt} \ \left( \frac{dt}{d\tau} \right)^2 + 2g_{t\phi} \, \left( \frac{dt}{d\tau} \right) \left( \frac{d\phi}{d\tau} \right) + g_{\phi\phi} \left( \frac{d\phi}{d\tau} \right)^2 + g_{rr} \left( \frac{dr}{d\tau} \right)^2 + g_{\theta\theta} \left( \frac{d\theta}{d\tau} \right)^2.
    \label{eq:B1}
\end{align}

The absence of explicit dependence on $t$ and $\phi$ in the metric components implies the existence of two conserved quantities $E$ and $L$ as discussed in Sec. \ref{sec:veffmagkerr}.

We now consider motion restricted to circular orbits in the equatorial plane, i.e., $\theta = \pi/2$, with $\dot{r} = \dot{\theta} = 0$. In this case, the orbital angular velocity (also referred to as the ``Keplerian frequency") is defined as \cite{CCGMmonople}

\begin{align}
    \Omega_\phi = \frac{-\partial_r g_{t\phi} \pm \sqrt{(\partial_r g_{t\phi})^2 - \partial_r g_{tt} \, \partial_r g_{\phi\phi}}}{\partial_r g_{\phi\phi}}\Bigg|_{\theta = \pi/2, , r = \text{const}}.
\end{align}

This expression follows directly from the geodesic equations under the conditions of circular motion, namely $dr/d\tau = 0$ and $d\theta/d\tau = 0$. These same conditions, when substituted into Eq.\eqref{eq:B1}, along with the relation $d\phi/d\tau = \Omega_\phi \ dt/d\tau$, allow us to solve for $dt/d\tau$. Substituting the result into Eq.\eqref{eq:metricEL1}, we obtain \cite{CCGMmonople}

\begin{align}
    E &= \frac{-g_{tt} - g_{t\phi} \Omega_\phi}{\sqrt{-g_{tt}-2g_{t\phi} \Omega_\phi -g_{\phi \phi} \Omega^2_\phi}}. \label{eq:Emetric}
\end{align}

Similarly, substituting into Eq.~\eqref{eq:metricEL2}, one obtains \cite{CCGMmonople}

\begin{align}
    L&= \frac{g_{t\phi} + g_{\phi\phi} \Omega_\phi}{\sqrt{-g_{tt}-2g_{t\phi} \Omega_\phi -g_{\phi \phi} \Omega^2_\phi}}. \label{eq:Lmetric}
\end{align}

In this spacetime, the specific angular momentum (also referred to as the proper angular momentum) of the test particle is defined as \cite{CCGMmonople}

\begin{align}
l \equiv -\frac{g_{t\phi} + g_{\phi\phi} \Omega_\phi}{g_{tt} + g_{t\phi}\Omega_\phi}. \label{eq:l_def}
\end{align}

When a particle on a stable circular orbit is perturbed, it undergoes small harmonic oscillations in the radial and vertical directions with frequencies $\Omega_r$ and $\Omega_\theta$, respectively. These are known as the radial and vertical epicyclic frequencies. The perturbations are introduced as:
\begin{align}
    r(t) &= r + \delta r(t), \\
    \theta(t) &= \frac{\pi}{2} + \delta\theta(t),
\end{align}
which correspond to oscillations about the equatorial plane $\theta=\pi/2$. Following Ref.\cite{Ryan1995}, the epicyclic frequencies are given by \cite{CCGMmonople}:

\begin{align}
    \Omega_r = \frac{g_{tt}+\Omega_\phi g_{t\phi}}{\sqrt{2 g_{rr}}} \bigg[\partial_r^2 \left( \frac{g_{\phi\phi}}{Y} \right) + 2l\partial_r^2  \left( \frac{g_{t\phi}}{Y} \right) + l^2 \partial_r^2 \left( \frac{g_{tt}}{Y} \right) \bigg]^{1/2} \Bigg|_{\theta = \pi/2, , r = \text{const}},
\end{align}

and

\begin{align}
    \Omega_\theta = \frac{g_{tt}+\Omega_\phi g_{t\phi}}{\sqrt{2 g_{\theta\theta}}} \bigg[\partial_\theta^2 \left( \frac{g_{\phi\phi}}{Y} \right) + 2l\partial_\theta^2  \left( \frac{g_{t\phi}}{Y} \right) + l^2 \partial_\theta^2 \left( \frac{g_{tt}}{Y} \right) \bigg]^{1/2} \Bigg|_{\theta = \pi/2, , r = \text{const}},
\end{align}

where

\begin{align}
    Y \equiv g_{tt}g_{\phi\phi} - g_{t\phi}^2.
\end{align}

Finally, two additional derived frequencies of significant astrophysical interest are the periastron and nodal precession frequencies, which arise from the difference between the orbital and epicyclic motions. These are defined as \cite{belloni2014fast}:

\begin{align}
    \Omega_\text{per} &= \Omega_\phi - \Omega_r, \label{Appendixper} \\
    \Omega_\text{nod} &= \Omega_\phi - \Omega_\theta. \label{Appendixnod}
\end{align}

These expressions are applicable to any stationary and axisymmetric geometry, such as the Kerr, magnetized Kerr, or Melvin spacetimes. In the main text, we apply these formulas to the magnetized Kerr background in Sec.~\ref{sec:Fundamental Frequinces}.

\bibliography{ref}
\bibliographystyle{apsrev4-2}

\end{document}